\documentclass[%
 reprint,
 superscriptaddress,
 amsmath,amssymb,
 aps,
]{revtex4-2}

\usepackage{graphicx}
\usepackage{dcolumn}
\usepackage{bm}
\usepackage{hyperref}
\usepackage[mathlines]{lineno}
\usepackage{siunitx}

\usepackage{custom}

\begin{document}

\preprint{APS/123-QED}

\title{Finite density QCD equation of state: critical point and lattice-based \texorpdfstring{$T'$}{T'}-expansion}

\author{Micheal Kahangirwe}\email{mkahangi@Central.uh.edu}
\affiliation{
 Department of Physics, University of Houston, Houston, TX 77204, USA
}

\author{Steffen A. Bass}
\affiliation{
 Department of Physics, Duke University, Durham, NC 27708, USA
}

\author{Elena Bratkovskaya}
\affiliation{%
 Helmholtz Research Academy Hesse for FAIR (HFHF), 
 GSI Helmholtz Center for Heavy Ion Physics, Campus Frankfurt, 60438 Frankfurt, Germany
}
\affiliation{%
 Institut für Theoretische Physik, Johann Wolfgang Goethe-Universität, Max-von-Laue-Str. 1, D-60438 Frankfurt am Main, Germany
}
\affiliation{
 GSI Helmholtzzentrum für Schwerionenforschung GmbH,Planckstrasse 1, D-64291 Darmstadt, Germany
}%

\author{Johannes Jahan}
\affiliation{
 Department of Physics, University of Houston, Houston, TX 77204, USA
}

\author{Pierre Moreau}
\affiliation{
 Department of Physics, Duke University, Durham, NC 27708, USA
}

\author{\\Paolo Parotto}
\affiliation{Pennsylvania State University, Department of Physics, State College, PA 16801, USA}
\affiliation{Dipartimento di Fisica, Universit\`a di Torino and INFN Torino, Via P. Giuria 1, I-10125 Torino, Italy}

\author{Damien Price}
\affiliation{
 Department of Physics, University of Houston, Houston, TX 77204, USA
}
\author{Claudia Ratti}
\affiliation{
 Department of Physics, University of Houston, Houston, TX 77204, USA
}

\author{Olga Soloveva}
\affiliation{%
 Helmholtz Research Academy Hesse for FAIR (HFHF), 
 GSI Helmholtz Center for Heavy Ion Physics, Campus Frankfurt, 60438 Frankfurt, Germany
}
\affiliation{%
 Institut für Theoretische Physik, Johann Wolfgang Goethe-Universität, Max-von-Laue-Str. 1, D-60438 Frankfurt am Main, Germany
}

\author{Mikhail Stephanov}

\affiliation{
Physics Department and Laboratory for Quantum Theory at the Extremes, University of Illinois at Chicago, Chicago, IL 60607, USA
}%
\affiliation{Kadanoff  Center  for  Theoretical  Physics,  University  of  Chicago,  Chicago,  Illinois  60637,  USA}

\date{\today}

\begin{abstract}

We present a novel construction of the QCD equation of state (EoS) at finite baryon density. Our work combines a recently proposed resummation scheme for lattice QCD results with the universal critical behavior at the QCD critical point. This allows us to obtain a family of equations of state in the range $0 \leq \mu_B \leq 700$ MeV and 25 MeV $\leq T \leq 800$ MeV, which match lattice QCD results near $\mu_B=0$ while featuring a critical point in the 3D Ising model universality class.
The position of the critical point can be chosen within the range accessible to beam-energy scan heavy-ion collision experiments. The strength of the singularity and the shape of the critical region are parameterized using a standard parameter set. 
We impose stability and causality constraints and discuss the available ranges of critical point parameter choices, finding that they extend beyond earlier parametric QCD EoS proposals. We present thermodynamic observables, including baryon density, pressure, entropy density, energy density, baryon susceptibility and speed of sound, 
that cover a wide range in the QCD phase diagram relevant for experimental exploration. 

\end{abstract}

\maketitle


\section{\label{sec:level1}Introduction\protect\\ }
The determination of the multi-dimensional QCD phase diagram is one of the main ingredients in understanding matter under extreme conditions of temperature and density, such as those created in heavy-ion collision experiments taking place at the Relativistic Heavy Ion Collider (RHIC) and the Large Hadron Collider (LHC). In nature, this kind of matter could be present in the core of neutron stars, and also in a primordial phase that permeated the universe a few microseconds after the Big Bang. To determine the QCD phase diagram we need to explore the thermodynamic behavior of strongly interacting matter, including its phase structure, equation of state (EoS) and critical phenomena \cite{Stephanov:2004wx}. 

In its most common representation \cite{Ratti:2022qgf,MUSES:2023hyz}, which involves temperature and baryon chemical potential or baryon density, the EoS at low net-baryon density is well understood. It exhibits a smooth crossover from a hadron gas to a quark-gluon plasma \cite{Aoki:2006we, Borsanyi:2010cj,Bellwied:2015rza,Bazavov:2011nk,HotQCD:2018pds} with a pseudo-critical temperature of $T_0 = 158.0\pm0.6$ MeV \cite{Borsanyi:2020fev}, and can be determined from first principles through lattice QCD simulations \cite{Borsanyi:2010cj,Borsanyi:2013bia,HotQCD:2014kol,Borsanyi:2018grb,Borsanyi:2021sxv,Borsanyi:2022qlh,Bollweg:2022fqq}. 
Several QCD models predict that the smooth crossover can turn into a first-order phase transition at high densities, thus implying the existence of a critical point on the QCD phase diagram \cite{Berges:1998rc,Halasz:1998qr,Stephanov:1998dy,Bzdak:2019pkr}. The search for the critical point is at the core of the Beam Energy Scan II (BESII) at RHIC, which completed data taking recently. The role of theorists in this program is to provide crucial tools to simulate and interpret the data. The equation of state is one of the fundamental quantities needed in the hydrodynamic description of the heavy-ion collision evolution.

Lattice simulations at finite chemical potential face challenges because of the fermion sign problem \cite{Splittorff:2007ck,Hsu:2010zza,Aarts:2012yal,Nagata:2021ugx}, which renders  traditional numerical techniques prohibitively costly. Despite recent developments in methods to directly simulate at finite chemical potentials, such as reweighting \cite{Giordano:2020roi,Borsanyi:2021hbk,Borsanyi:2022soo}, these are still limited to small volumes and rather coarse lattices. This has so far prevented realistic direct simulations in the most intriguing region of the QCD phase diagram. Therefore, the expected first-order phase transition from hadron gas to quark-gluon plasma at high density, as well as the critical point terminating this transition \cite{Stephanov:1998dy,Stephanov:2004wx} are still out of the reach of lattice simulations. Extrapolation techniques, such as Taylor expansion \cite{Gavai:2003mf,Allton:2005gk,Borsanyi:2012cr,
Bellwied:2015lba,Ding:2015fca,Bazavov:2017dus,HotQCD:2018pds,
Bazavov:2020bjn}, analytic continuation from imaginary chemical potential \cite{deForcrand:2002hgr,DElia:2002tig,DElia:2009pdy,
Bonati:2015bha,
Bellwied:2015rza,DElia:2016jqh,Bonati:2018nut,Borsanyi:2018grb,Borsanyi:2020fev} and Pad\'e approximation \cite{Bollweg:2022rps,Bollweg:2022fqq}, are usually employed to extend lattice QCD thermodynamic results to finite densities. However, they are limited in their applicability to small chemical potentials. 

An important tool for the theoretical interpretation of experimental results are hydrodynamic simulations 

\cite{Jeon:2015dfa,Luzum:2008cw,Gale:2012rq,Karpenko:2015xea,Niemi:2015bpj,Carzon:2023zfp,Aguiar:2007zz,Dore:2020jye}, which describe the evolution of the fireball produced in heavy-ion collisions. Although modifications to the relativistic viscous hydrodynamic approach are required close to the critical point \cite{Jeon:2015dfa,Stephanov:2017ghc}, it is crucial that the equation of state (EoS) used in these simulations encompasses all existing theoretical knowledge and accurately represents the singularity related to the QCD critical point in a predetermined and adjustable way. Moreover, the EoS as well as the properties of partons and their interactions are probed directly within microscopic transport approaches, wherein partonic and hadronic degrees of freedom are propagated explicitly \cite{Moreau:2019vhw}.

In an attempt to provide a tool to address these issues, the BEST collaboration developed a family of equations of state, based on the lattice QCD Taylor expansion, with a 3D Ising model critical point which matches lattice results at low chemical potential  \cite{Parotto:2018pwx,Mroczek:2020rpm,An:2021wof,Karthein:2021nxe}. However, this approach is limited to $\mu_B\leq450$ MeV, because unphysical oscillation inherited from the Taylor expansion appear in some observables at large $\mu_B$ \cite{Ratti:2007jf,Parotto:2018pwx}.

It is important to note that the temperature of the hypothetical chiral critical  point should not exceed the critical temperature of the chiral phase transition (for $m_u=m_d=0$) $T^{0}_c = 132^{+3}_{-6}$ MeV \cite{Karsch:2019mbv, HotQCD:2019xnw}. Lattice QCD simulations disfavor the existence of the critical point at $\mu_B\leq300$ MeV \cite{Borsanyi:2020fev}. Besides, several recent results seem to converge in predicting a critical point location at $560\leq\mu_B\leq650$ MeV \cite{Fu:2019hdw,Gao_2020,Gao_2021,Gunkel_2021,Hippert:2023bel}. For this reason, and to properly support the BESII at RHIC that can cover a range up to $\mu_B\lesssim700$ MeV, the BEST collaboration EoS needs to be extended to larger values of $\mu_B$. While some results exist in the literature \cite{Kapusta:2021oco}, where a critical scaling function was developed on top of an EoS with a smooth crossover between hadrons and quarks, here we follow the same strategy as the BEST collaboration EoS: we introduce the 3D Ising critical point into a lattice-QCD-based EoS. However, instead of using the Taylor expansion method, we build our EoS on the basis of the new expansion scheme developed in \cite{Borsanyi:2021sxv, Borsanyi:2022qlh}. This will allow us to reach a value of chemical potential $\mu_B\sim700$ MeV.

The manuscript is organized as follows. In section \ref{sec:lattice_EoS} we recall the lattice QCD approaches: Taylor expansion and alternative $T$-expansion scheme developed by the Wuppertal-Budapest lattice QCD collaboration in \cite{Borsanyi:2021sxv, Borsanyi:2022qlh}. Section \ref{subsec:mapping} focuses on the mapping of the 3D Ising model onto the QCD coordinates. Moving on to section \ref{sec:merging}, we discuss the merging of the lattice QCD equation of state with the critical one. In section \ref{sec:Thermodynamics}, we present the thermodynamic quantities with a critical point, and in section \ref{sec:Constraint} we explore the constraints on the parameter space.  Conclusions and outlook will be provided in section \ref{sec:conclusion}. Finally, in Appendix \ref{appendixA},\ref{appendixB} and \ref{appendixC} we provide detailed derivation for the formulas used.
The code that generates the family of equations of state presented in this paper can be downloaded from \cite{code}.

\section{\label{sec:lattice_EoS}Lattice Equation of State}

\subsection{\label{subsec:Taylor}Taylor Expansion}
Taylor expansion is the most straightforward way to extend the equation of state to finite $\mu_B$. It consists of a sum of all pressure derivatives (susceptibilities), computed on the lattice at $\mu_B=0$, multiplied by powers of a dimensionless expansion parameter 
$\left(\frac{\mu_B}{T}\right)$. Because of charge conjugation symmetry, only even susceptibilities contribute 

\begin{equation}
    \frac{P(T,\mu_B)}{T^4} = \sum_{n=0}{\frac{1}{2n!}\chi_{2n}^B(T,\mu_B = 0)\left(\frac{\mu_B}{T}\right)^{2n}} \, \, ,\label{PressureTaylor}
\end{equation}
where the coefficients are:
 \begin{equation*}
    \chi_n^B(T) = \left(\frac{\partial^n}{\partial(\mu_B/T)^n} \frac{P(T,\mu_B)}{T^4} \right)_{\mu_B = 0} \, \, .
\end{equation*} 

In this paper, we will focus on the baryon density, which is defined as the first derivative of the pressure with respect to $\mu_B$:  

\begin{eqnarray}
    \frac{n_B(T,\mu_B)}{T^3} &=& \frac{\partial}{\partial(\mu_B/T)}\frac{P(T,\mu_B)}{T^4}
    \nonumber\\
    &=&\sum_{n=1}^\infty \frac{1}{(2n-1)!}\chi_{2n}^B(T)\left(\frac{\mu_B}{T}\right)^{2n-1}.\label{BaryonTaylor}
\end{eqnarray}
To completely evaluate the baryon density in Eq. (\ref{BaryonTaylor}), we would need all the coefficients computed on the lattice, which are not readily available 
due to limitations in computational power. Currently, coefficients are available at finite lattice spacing up to order $\mathcal{O}\left(\frac{\mu_B}{T}\right)^6$ \cite{HotQCD:2017qwq,Guenther:2017hnx} and even $\mathcal{O}\left(\frac{\mu_B}{T}\right)^8$ \cite{Borsanyi:2018grb,Bollweg:2022fqq}, and in the continuum limit in a smaller volume \cite{Borsanyi:2023wno}, which leads to the following limitations of the method:

\begin{itemize}
    \item The chemical potential range is limited to $\frac{\mu_B}{T} < 3$, despite  large computational power \cite{Bollweg:2022rps,Bollweg:2022fqq}.
    \item At large $\mu_B/T$, some observables exhibit unphysical, ``wiggly'' behavior due to the truncation of the Taylor series \cite{Ratti:2007jf,Parotto:2018pwx}.
    \item The inclusion of an additional higher-order term does not improve this behavior.
    \item The Taylor expansion struggles to account for a transition temperature that depends on the chemical potential, since it is performed at constant temperature. In \cite{Brandt:2018omg}, the Taylor expansion was tested for finite isospin chemical potential by comparing it to the direct lattice simulation, and a breakdown was observed at the critical chemical potential.

\end{itemize}

The above limitations make it difficult to model and constrain the existence of the critical point if it is located at high density.
In \cite{Parotto:2018pwx,Mroczek:2020rpm,An:2021wof,Karthein:2021nxe}, the BEST collaboration exploited the universality class of the 3D Ising model to introduce a critical point into the equation of state by separating the free energy density into a critical contribution and a non-critical one, so that the sum of the Taylor expansion coefficients up to $\mathcal{O} ( (\mu_B/T)^4)$ reproduces the lattice results. For the baryon density, this procedure works as follows
\small
\begin{align}\nonumber
    n_B(T,\mu_B) &= T^3\sum_{n=1}^2\frac{1}{(2n-1)!}\chi_{2n}^{B ~\rm non-Ising}(T)\left(\frac{\mu_B}{T}\right)^{2n-1} \\
    &~~~~~~~~~~~~~~~~~~~~~~~~~~~~ + \frac{T_C^4}{T} n_{B}^{\rm Ising}(T,\mu_B)
\end{align}
where $n_{B}^{\rm Ising}$ is the contribution to the baryon density with the singular behavior appropriate for the 3D Ising critical point, and the coefficients $\chi_{n}^{B~\rm non-Ising}(T)$ satisfy
\begin{equation*}
    \chi_n^{B~\rm lat}(T) = \chi_n^{B~\rm non-Ising}(T) + \frac{T_C^4}{T^4}\chi_n^{B~\rm Ising}(T)
\end{equation*}
for $n=0,~2,~4$, where $\chi_n^{B~\rm lat}(T)$ are the input from lattice QCD and $\chi_n^{B~\rm Ising}(T)$ represent the critical contribution to the expansion coefficients.
Although this construction works well, it was observed that at large values of $\mu_B$,  wiggles appear in the thermodynamic observables, particularly the baryon density and speed of sound, for some parameter choices. This is due to the truncation in the Taylor expansion of the non-Ising contribution to the observables, which limits the current applicability of this equation of state.

\subsection{\label{sec:alternative}\texorpdfstring{$T'$}{T'}-Expansion Scheme}

\begin{figure}[!hbtp]
    \centering
\includegraphics[scale=0.235]{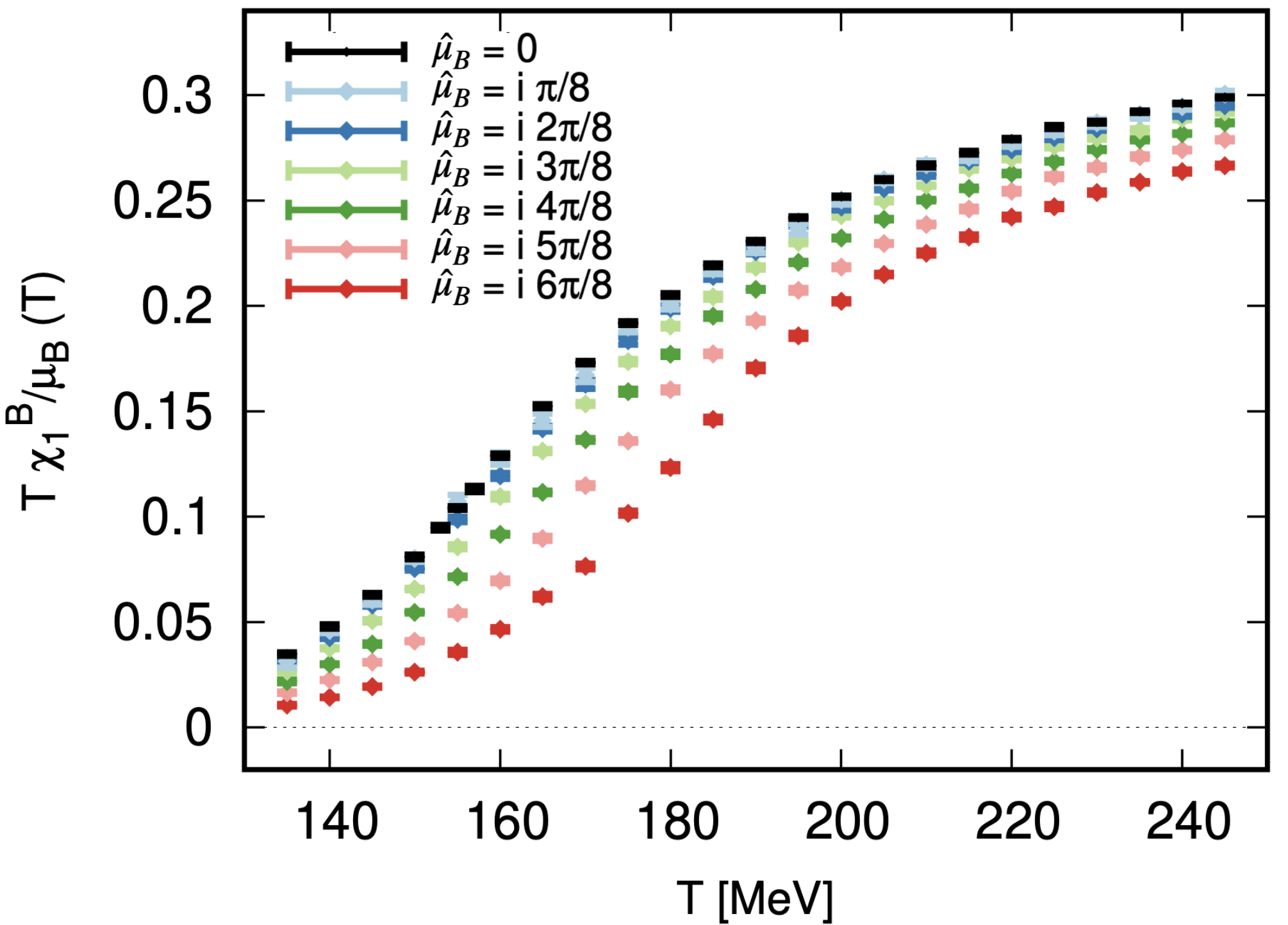}
\includegraphics[scale=0.19]{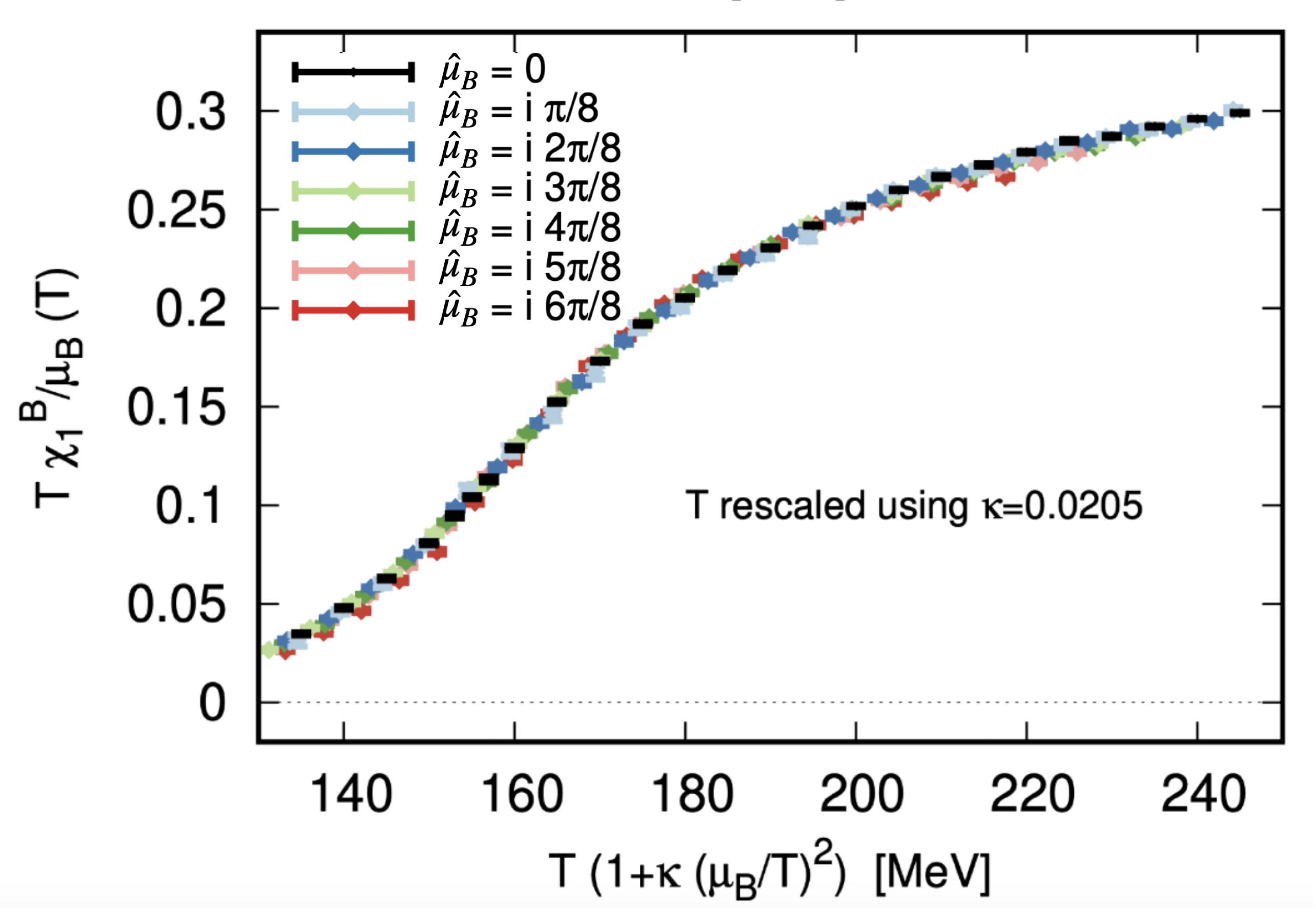}
    \caption{
    Upper panel: scaled baryon density ${\chi_1^B(T,\mu_B)}/{\hat\mu_B}$, as a function of  temperature for different values of scaled imaginary baryon chemical potential $\hat{\mu}_B \equiv {\mu_B}/{T}$ (labeled using different colors).
    Lower panel: the same quantity, but with the temperature rescaled by a factor $1+\kappa\hat\mu_B^2$, with $\kappa = 0.0205$. In terms of the rescaled temperature the curves representing different $\hat\mu_B$ collapse onto the same curve.
    The points labeled $\hat\mu_B=0$ correspond to the limit $\mu_B\to0$ which is  the baryon number susceptibility 
    $\chi^B_2(T,0)$
    (The figure is taken from  Ref.~\cite{Borsanyi:2021sxv}).
    } 
    \label{fig:altExS}
   \end{figure}

To address some of the limitations of the Taylor expansion outlined above, the Wuppertal-Budapest lattice QCD collaboration developed a novel resummation scheme, which can reach higher values of chemical potential and handle the QCD transition line  \cite{Borsanyi:2021sxv, Borsanyi:2022qlh}. 
The scheme is based on the observation \cite{Borsanyi:2021sxv} that the crossover in terms of the scaled baryon density $T\chi_1^B/\mu_B$ as a function of $T$ looks very similar at different (imaginary) values of scaled chemical potential $\mu_B/T$, with most of the difference being a $\mu_B$-dependent shift of $T$ -- see Fig.\ref{fig:altExS}.

This observation can be formalized by expressing baryon density $n_B\equiv\chi_1^B T^3$ in the form

\begin{equation}
T\frac{\chi_1^B(T,\mu_B)}{\mu_B} = \chi_2^B(T',0) \label{Eq:alternativeExs}
\end{equation}
which {\em defines} the ``rescaled temperature" $T'(T,\mu_B)$. At $\mu_B=0$, $T'$ is the same as $T$. At non-zero $\mu_B$ function $T'(T,\mu_B)$ is such that the crossover in terms of $T{\chi_1^B(T,\mu_B)}/{\mu_B}$ occurs at the same $T'$ and has the same shape. The function $T'$ can be then expanded in powers of $\mu_B/T$ at fixed $T$:
\begin{equation}
    T'(T,\mu_B) = T\left[1 + \kappa_2^{BB}(T)\left(\frac{\mu_B}{T}\right)^2 + \kappa_4^{BB}(T)\left(\frac{\mu_B}{T}\right)^4 + ... \right] \, \label{Shifting}
\end{equation}
where the Taylor expansion coefficients $\kappa_2^{BB}$, etc. are almost constant as functions of $T$ in the transition region, while the rapid changes in EoS associated with the crossover are mostly captured by the function $\chi_2^B(T',0)$ (see Fig.\ref{parametrized} below.).

The ``$T'$-expansion" scheme is essentially a re-shuffling of the Taylor expansion in Eq.\eqref{BaryonTaylor}, and the coefficients $\kappa_n^{BB}(T)$ can be expressed in terms of the susceptibilities $\chi_{2n}^B(T)$:
\begin{align}
    \kappa_2^{BB}(T) &= \frac{1}{6T} \frac{\chi_4^B(T)}{{\chi_2^B}'(T)} \label{kappa2f}\\
   \kappa_4^{BB}(T) &= \frac{1}{130 T{\chi_2^B}'(T)^3} \left(3{\chi_2^B}'(T)^2\chi_6^B(T) - 5{\chi_5^B}''\chi_4^B(T)^4\right).
   \nonumber
\end{align}
These coefficients were obtained in high-statistics lattice QCD simulations  \cite{Borsanyi:2021sxv}. As expected, compared to the sharply rising $\chi_2^B(T)$, $\kappa_2^{BB}$ shows a very mild temperature dependence around the transition region, which makes the $T'$-expansion scheme more favorable than the Taylor expansion since it does not introduce the wiggly behavior in the EoS at large $\mu_B$. Moreover, the fact that $\kappa_4^{BB}$ is shown in \cite{Borsanyi:2021sxv} to be consistent with zero hints at a faster convergence compared to the Taylor series. 

These results agree with the one used in \cite{Bazavov:2017dus} for ``lines of constant physics" calculated up to $\mathcal{O}(\mu_B^4)$. 
As suggested in \cite{Borsanyi:2021sxv}, as long as $\chi_1^B/\hat{\mu}_B$ is a monotonic function of $T$, the finite-density physics can be encoded into the $T'(T,\mu_B)$ function. As a result, we can embed the singularity associated with the critical point and the first-order phase transition into $T'(T,\mu_B)$, as we will show in Section \ref{sec:merging}.

\subsection{\label{sec:lattice}Lattice data}
Lattice results for the susceptibility $\chi_2^{B}(T,0) \equiv \chi_2^{B}(T)$ and coefficients 
$\kappa_2^{BB}(T)$ are available only over a limited range of (discrete) temperatures. To obtain a smooth description of the equation of state in the temperature range $25 \, \text{MeV}  \leq T \leq 800$ MeV,  we first merge the lattice results at finite temperature and $\mu_B=0$ with 2+1 flavors and physical quark masses from the Wuppertal-Budapest Collaboration \cite{Borsanyi:2010cj,Borsanyi:2013bia,Bellwied:2015lba,Borsanyi:2018grb}  with the hadron resonance gas (HRG) model results \cite{Vovchenko:2019pjl}, which provide a good description of the thermodynamics up to $T=120 $ MeV,  using the most up to date particle list (list PDG2021+) \cite{,Zyla:2020zbs,SanMartin:2023zhv}. We then fit these results to cover a large range of temperatures.

\begin{figure}[!htbp]
    \centering
    \includegraphics[scale=0.27]{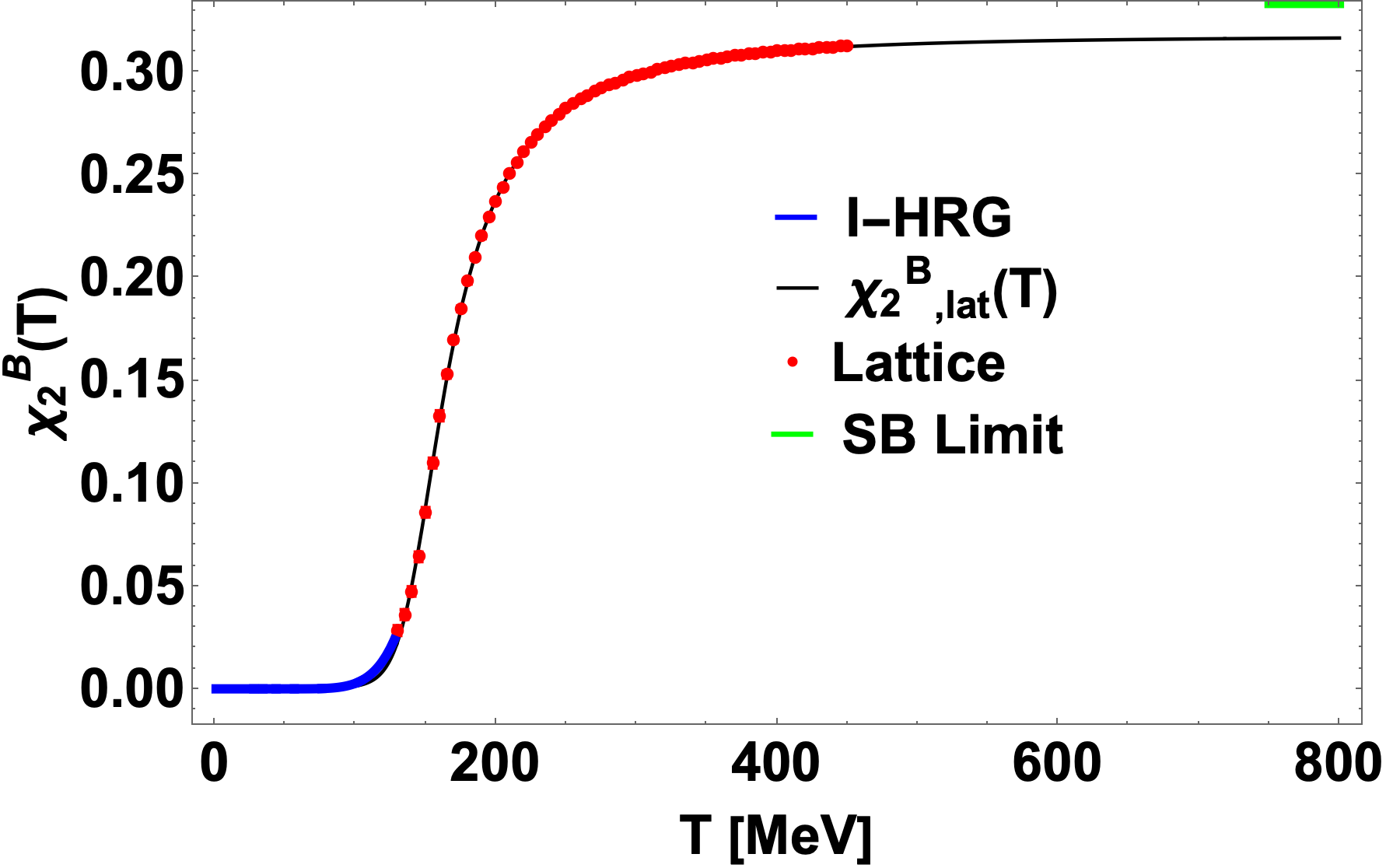}
    
    \includegraphics[scale=0.43]{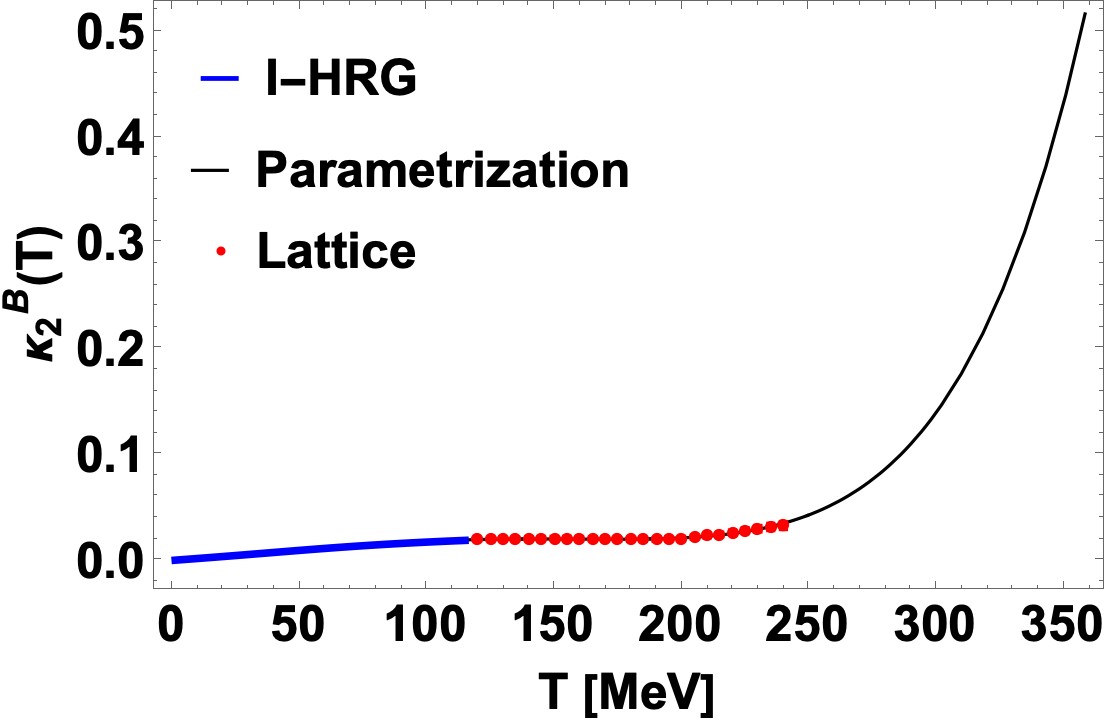}
    \caption{
 Top panel: parameterized baryon susceptibility $\chi^{B}_{2,\rm lat} (T)$ (black curve), in the range $25 \, \text{MeV} <T<800 \text{MeV}$. The Stefan-Boltzmann (SB) limit value is shown in green. Bottom panel: parametrized alternative expansion coefficient as a function of the temperature. In both panels, the solid blue curve corresponds to the hadron resonance gas (HRG) model prediction, while the red dots represent continuum extrapolated lattice QCD results.}
    \label{parametrized}
\end{figure}

For convenience we introduce an auxiliary variable 
$x = {T}/({200 \, \text{MeV}})$.
For $\chi_{2,\text{lat}}^B(T)$, we employ four free parameters $d_i$, such that the crossover occurs at $x \approx d_1$, and its width $\Delta x \sim d_1/d_2$ is controlled by $d_2 \gg 1$, while $d_3\left(1-\frac{d_4^2}{x^2}\right)$ provides large-$x$ asymptotics:
\begin{equation}
   \chi_{2,\rm lat}^B(T) = \left(\frac{2m_p}{\pi x}\right)^{3/2} \frac{e^{-m_p/x}}{1+\left(\frac{x}{d_1}\right)^{d_2}} + d_3\frac{e^{-d_4^2/x^2 - d_5^4/x^4}}{1+ \left(\frac{x}{d_1}\right)^{-d_2}}
    \label{Chi2Eq}
\end{equation}
where, $m_p\approx4.7$ denotes the proton mass (in units of 200 MeV). The first term, typically very small, yields the correct low-temperature asymptotics for $\chi_2^B$ in QCD, representing the nonrelativistic contribution of nucleons/antinucleons. Best-fit coefficients for $\chi_{2,\text{lat}}^B(T)$ are listed in Table \ref{chi_2_parameters}, and the resulting parametrization is shown in the top panel of Fig.~\ref{parametrized}.

\begin{table}[!tbh]
    \centering
    \begin{tabular}{|c|c|c|c|c|}
        \hline
        $d_1$ & $d_2$ & $d_3$ & $d_4$ & $d_5$ \\
        \hline
        $0.73$ & $11.19$ & $0.32$ & $0.20$ & $0.69$ \\
        \hline
    \end{tabular}
    \caption{Coefficients of the parameterized $\chi_{2,\text{lat}}^B(T)$ in Eq.~(\ref{Chi2Eq}) for $25~\text{MeV} \leq T \leq 800~\text{MeV}$}
    \label{chi_2_parameters}
\end{table}

Because $\kappa_{2}^{BB}$, unlike $\chi_2^B$, varies slowly in the temperature region of our interest, we can use a rational fit. We enforce the expected small-temperature linear behaviour which follows from the dominant exponential behavior $\chi^B_4,\chi^B_2\sim e^{-m_p/x}$ and Eq.~\ref{kappa2f}: $\kappa_2^{BB}/x\to A_1=1/(6m_p)\approx 0.035 $ Similarly, the leading large-temperature behavior follows from Eq.\eqref{Chi2Eq}, i.e., $\chi_2=d_3-d_3d_4^2/x^2+\dots$ and $\chi_4=2/({9\pi^2})+\dots$. Eq.~\ref{kappa2f} then gives 
$\kappa_{2}^{BB}/x^2\to A_2= 1/(54 d_3 d_4^2) \approx 1.47 $. We then use the following fitting function:
\begin{equation}\label{kappaEq}
    \kappa_{2}^{BB}(T) = \frac{A_1 b_0 x + a_2 x^2 +a_3 x^3 +A_2 x^4 }{b_0 + b_1 x +x^2  } 
\end{equation}
where, again, $x = {T}/({200 \, \rm MeV})$.  Best-fit parameters for $\kappa_2^{BB}(T)$ are listed in Table \ref{Kappa_parameters}, and the resulting parametrization is shown in the bottom panel of Fig.~\ref{parametrized}.

\begin{table}[!tbh]
    \centering
\begin{tabular}{|c|c|c|c|c|c|c|c|}
      \hline
      $a_2$ & $a_3$ & $b_0$ & $b_1$    \\
      \hline
     0.652 & -2.60 & 21.4& -9.81  \\
      \hline
\end{tabular}
 \caption{Coefficients of the rational parameterization for $\kappa^{BB}_{2}(T)$ in Eq (\ref{kappaEq}) for $25 ~\text{MeV}  \leq T \leq 800 ~\text{MeV}$.}  

    \label{Kappa_parameters}
\end{table}

\section{\label{subsec:mapping}Mapping the 3D Ising model to QCD}

Close to the critical point, the correlation length of a thermodynamic system diverges, making microscopic (short-distance) features irrelevant. Consequently, systems with similar global symmetries exhibit similar, universal behavior, even though they may differ in their microscopic degrees of freedom. Well-known examples of this phenomenon include liquid-gas and ferromagnetism, which share critical exponents within the same universality class as the 3D Ising model \cite{fisher1974critical, Pradeep:2019ccv}.
The critical point of Quantum Chromodynamics (QCD), if it exists, also belongs to the 3D Ising model universality class \cite{Rajagopal:1992qz}. Hence, its critical behavior is characterized by the same critical exponents, which describe the scaling of physical quantities in the thermodynamic variables near the critical point \cite{Rajagopal:1992qz}. 

\subsection{Scaling: 3D Ising Model }

In this work, we employ the same form of the scaling equation of state as used in the BEST collaboration equation of state. 
The parameterization of magnetization, denoted by $M$, reduced temperature ($r$) and external magnetic field ($h$) in terms of additional scaling standard  variables $R$ and $\theta$, is given as follows \cite{Parotto:2018pwx, Guida:1996ep, Caselle:2020tjz,Nonaka:2004pg,Kapusta:2022pny,Schofield:1969zza,Lu:2023msn}:
\begin{eqnarray}
    M = ~&& M_0R^\beta\theta\\
    h = ~&& h_0R^{\beta\delta}\tilde{h}(\theta)\\
    r = ~&& R(1-\theta^2).
\end{eqnarray}

The scale invariant ``angular" variable $\theta$ describes the position of a point on $r,h$ plane relative to the $h=0$ ($\theta=0$) and $r=0$ ($\theta=1$) axes, allowing a non-singular description of both regimes for $R\neq 0$. The ``radial" variable $R$ measures the distance from the critical point, $R=0$.
The parameterization involves an odd function  $\tilde{h}(\theta) = \theta(1 + a \theta^2 + b \theta^4)$, where $a = -0.76201$ and $b = 0.00804$. The critical exponents for the 3D Ising model are $\beta = 0.326$ and $\delta = 4.80$. While $R$ is non-negative ($R \ge 0$), $|\theta|$ should not exceed the first non-trivial zero of $\tilde{h}(\theta)$, denoted as $\theta_0 \simeq 1.154$ and corresponding to $r<0$, $h=0$ axis.
To fix the values of the normalization constants $M_0$ and $h_0$, two conditions $M(r = -1, h = 0^+) = 1$ and $M(r = 0, h=1)=1$  are used. These conditions result in $M_0 = 0.605$ and $h_0 = 0.364$.
It is important to note that this parametric representation gives a non-globally invertible mapping from $(R, \theta) \mapsto (r, h)$. The critical point is located at $(r = 0, h = 0)$, and when $r < 0$, there is a smooth transition (crossover), while $r > 0$ corresponds to a first-order phase transition.

In this parameterization form, the pressure is defined in terms of the most singular part of the Ising Gibbs free energy $G(R,\theta)$:
\begin{eqnarray}
    G (R,\theta) = h_0 M_0 R^{2-\alpha}(\theta \tilde{h}(\theta) - g(\theta)) \;,
    \label{gibbs_pressure}
\end{eqnarray}
where
\begin{eqnarray*}
    g(\theta) =&&~  c_0 + c_1(1 - \theta^2) + c_2(1 - \theta^2)^2 + c_3(1-\theta^2)^3 \;, \\
    c_0 =&& \frac{\beta}{2-\alpha}(1 + a + b) \;, \\
    c_1 =&& -\frac{1}{2}\frac{1}{\alpha-1}((1-2\beta)(1 + a + b) -2\beta(a+2b) \;, \\
    c_2 =&& -\frac{1}{2\alpha}(2\beta b - (1- 2\beta)(a + 2b)) \;, \\
    c_3 =&& -\frac{1}{2(\alpha + 1)}b(1 - 2\beta) \;, 
\end{eqnarray*}
with $\alpha = 0.11$ another critical exponent, related to $\beta, \delta$ by the relation $2-\alpha = \beta (\delta + 1)$. 

\subsection{Mapping 3D Ising coordinates to QCD coordinates }

\begin{figure*}[!ht]
  \centering
  \includegraphics[width=\textwidth, height=10cm]{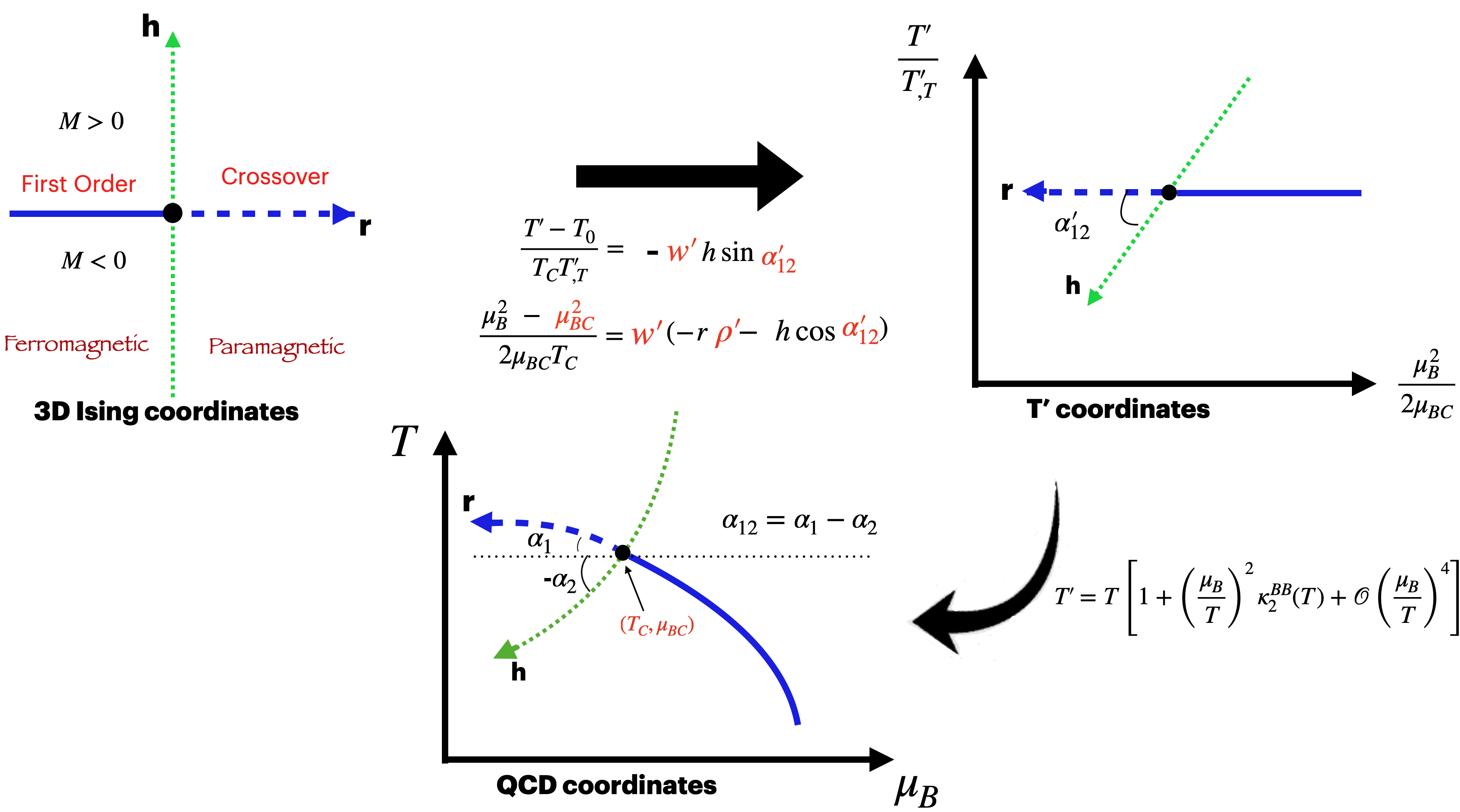}
  \caption{The top-left plot represents the 3D Ising model axes, with a critical point located at \((r=0, h=0)\). The top-right plot displays the $T'$-expansion scheme coordinates, with a critical point at \((T' = T_0, \mu_B = \mu_{BC})\). Finally, the bottom plot corresponds to the QCD coordinates, featuring a critical point located at \((T_C, \mu_{BC})\). The parameters in red $\mu_{BC}$, \(w'\),\(\ \rho'\) and \(\alpha'_{12}\) are the free parameters.}
  \label{fig:Mapping}
\end{figure*}

To map from the 3D Ising model to QCD, we employ a two-step non-universal mapping, as shown in Fig. \ref{fig:Mapping}. This process involves transforming the 3D Ising control parameters, namely the reduced temperature ($r$) and the external magnetic field ($h$), initially into the $T'$-expansion scheme coordinates represented by the "rescaled temperature" ($T'$) and the squared baryon chemical potential ($\mu_B^2$) using Eq. (\ref{quadrticmappingEq}) below. Subsequently, using the relation between $T$ and $T'$,  we map these coordinates to the QCD parameters, specifically the temperature ($T$) and the baryon chemical potential ($\mu_B$). 
 To ensure that the transition of the Ising model $(h=0)$ aligns with the QCD crossover line, we apply the following transformation:
\begin{eqnarray}\nonumber
    \frac{T' - T_0}{T_C T'_{,T}} & = & -w' h \sin\alpha'_{12} \\
    \frac{\mu_B^2 - \mu_{BC}^2}{2\mu_{BC}T_C} & = & w' (-r\rho' - h \cos\alpha'_{12})
    \label{quadrticmappingEq}
\end{eqnarray}
where $T_0$ is the transition temperature at $\mu_B=0$, $T_C$ and $\mu_{BC}$ are the temperature and chemical potential at the critical point, \(T'_{,T} \equiv (\partial T'/\partial T)_\mu\) at the critical point, and the free parameters \(w', \rho'\), and \(\alpha'_{12}\) act as scaling factors for variables \(r\) and \(h\). \(w'\) determines the size of the critical region, and \(\rho'\) modifies its shape. The scaling can also be accomplished by modifying the angle \(\alpha'_{12}\). These free parameters can easily be related to the ones used by the BEST Collaboration \cite{Parotto:2018pwx} in the linear mapping shown in Eq. \eqref{linearmapping}. By linearizing Eq. \eqref{quadrticmappingEq} around the critical point, and comparing to the coefficients of \(r\) and \(h\) in Eq. \eqref{linearmapping}, we obtain the following relations between \(w'\), \(\rho'\), \(\alpha'_{12}\) and \(w\), \(\rho\), \(\alpha_{1}\), \(\alpha_{2}\):
\begin{subequations}
\label{Eq:Tmap_BEST}
\begin{align}
\tan\alpha_{12}' &= \tan\alpha_1-\tan\alpha_2 \label{Eq:tan12} \\
    w' &= w\frac{1}{\cos\alpha_1}{\sqrt{(\cos\alpha_1\cos\alpha_2)^2+(\sin\alpha_{12})^2}} \label{Eq:w} \\
    \rho' &= \rho\frac{\cos^2\alpha_1}{\sqrt{(\cos\alpha_1\cos\alpha_2)^2+(\sin\alpha_{12})^2}} \label{Eq:rho}.
\end{align}
\end{subequations}

The parameters \((w, \rho)\) act as scaling factors for the variables \(r\) and \(h\), where \(w\) determines the size of the critical region, and \(\rho\) modifies its shape. The difference \(\alpha_{12}\) between  \(\alpha_{2}\) and  \(\alpha_{1}\) also controls the strength of the discontinuity. 
Equations~\eqref{Eq:Tmap_BEST} can be inverted to give:
\begin{subequations}
\begin{align}\label{Eq:alpha2}
    \tan\alpha_2 & = \tan\alpha_1 - \tan\alpha_{12}'; \\ 
    w & = w'\cos\alpha_{12}'\sqrt{1 + \left(\tan\alpha_1 - \tan\alpha_{12}'\right)^2}; \\
    \rho  & = \rho'\frac{1}{\cos\alpha_1\cos\alpha_{12}'\sqrt{1 + \left(\tan\alpha_1 - \tan\alpha_{12}'\right)^2}}.
\end{align}
\end{subequations}

A more concise way of converting from one set of parameters to another is as follows.
First, find $\alpha_{12}'$ from $\alpha_2$ using Eq. \eqref{Eq:alpha2}. Then use it to find $w'$ from
\begin{equation} \label{Eq:wpw}
    w'\cos\alpha_{12}' = w \cos\alpha_2\,.
\end{equation}
Then find $\rho'$ by solving
\begin{equation} \label{Eq:rhoprho}
    \rho'w' = \rho w \cos\alpha_1\,.
\end{equation}
It is also important to identify the parameters that control the strength of the discontinuity, which can be clearly seen in the expansion of the specific heat at constant pressure \(C_p\).
The leading singular behavior of \(C_p\) is given by:
\begin{align}
    C_p = T^3\left(\frac{(s_c/n_c)\sin\alpha_1 - \cos\alpha_1}{w\sin\alpha_{12}}\right)^2 G_{hh}\left(1+\mathcal{O}(r^{\beta\delta -1})\right)
\end{align}
in terms of the standard BEST collaboration parameters \cite{Parotto:2018pwx}, where \(G_{hh}\) is the order parameter susceptibility in the Ising model, while \(s_c\) and  \(n_c\) are the critical entropy and baryon density respectively. Since \(G_{hh}\) is the same for all mapping parameters, we can use the coefficient in front of it as a ``universal" measure of the strength of the singularity. It is then obvious that the strength measured that way depends on \(\alpha_{12}\) and \(w\) (at fixed \(\alpha_1\)) only via the combination \(w \sin\alpha_{12}\).

The mapping in Fig. \ref{fig:Mapping} comes with inherent advantages. The tunable free parameters can be guided by physics, such as the physical value of the quark masses, stability, and causality of the equation of state \cite{Pradeep:2019ccv}. This feature enables us to transport any physical quantity in 3D Ising to any point in the QCD phase diagram, and as the mapping is an even function in the baryon chemical potential, it ensures the expected charge conjugation symmetry.

\subsection{Transition Line}

With the mapping defined in Eq.(\ref{quadrticmappingEq}), the location of the transition line in the phase diagram is naturally determined. The transition line $T_C(\mu_{BC})$ is such that $T_C$ and $\mu_{BC}$ have to satisfy $T'(T,\mu_B) = T_0$, where $T_0$ is the crossover temperature at $\mu_B = 0$. For convenience, we use the pseudo-critical temperature related to chiral symmetry restoration $T_0 = 158$ MeV computed from the lattice in \cite{Borsanyi:2020fev}. In addition, for simplicity, we identify $T'$ with $T'_{\text{lat}}(T,\mu_B)$ in \cite{Borsanyi:2021sxv} up to second order in $\mu_B/T$. 
From Eq. \eqref{gibbs_pressure}, we make use of the mapping to express the critical pressure $P^{\rm crit}$ as a function of temperature and chemical potential:
\begin{eqnarray}
    P^{\rm crit}(T,\mu_B) = -T^4 G (R(T,\mu_B),\theta(T,\mu_B)) \, \, .
\end{eqnarray}
The critical baryon density is then defined as 
\begin{eqnarray}\label{eq:chi1-crit}
   \chi_1^{B~\rm crit} =  \frac{n_B^{\rm crit}(T,\mu_B)}{T^3} = \frac{\partial (P^{\rm crit}(T,\mu_B)/T^4)}{\partial (\mu_B/T)}\Big|_T \, \, .
\end{eqnarray}

With this mapping the critical point is also forced to sit on the transition line by construction. Therefore, the number of free parameters is reduced since the critical temperature follows from the choice of critical chemical potential, and the angle $\alpha_1$ is given by the slope of the transition line at the critical point:
 \begin{eqnarray}
     \alpha_1 = \tan^{-1}\left(\frac{2 \kappa_2(T_C)\mu_{BC}}{T_C T'_{,T}}\right) \;.
     \label{Eq:alpha1}
 \end{eqnarray}

\begin{figure*}[!htbp]
\includegraphics[width=\textwidth,height=10cm]{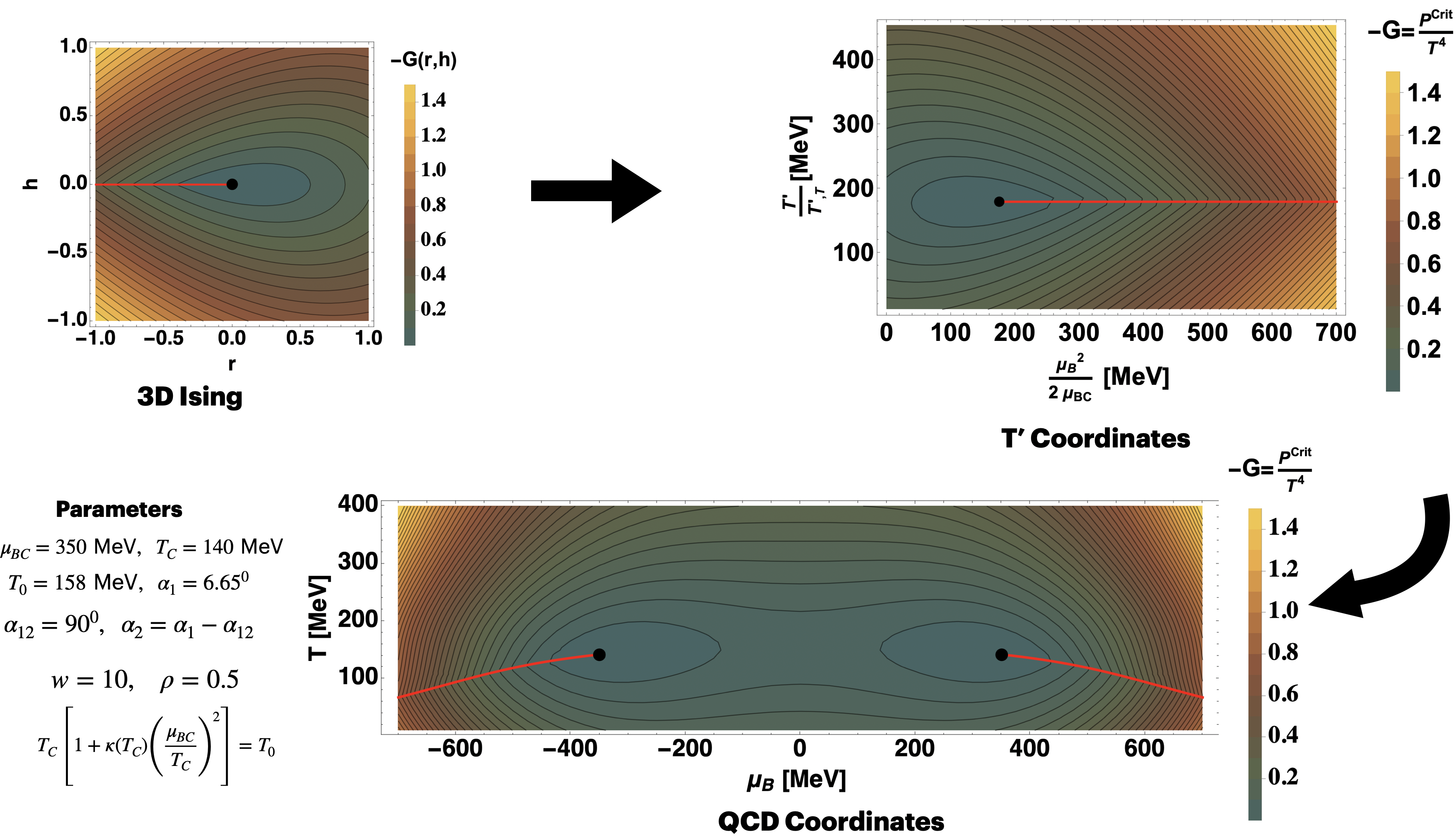}
  \caption{The figure comprises three contour plots of the critical (singular) contribution to pressure, in three different coordinate systems related to each other by transformations shown in Fig.\ref{fig:Mapping}. The top-left plot uses the Ising model coordinates $(h,~r)$, with the critical point located at $(0,0)$. The top-right plot uses coordinates $(T/T'_{,T},\mu_B^2/(2\mu_{BC}))$,  with the critical point at $(T_0/T'_{,T}, \mu_{BC}/2)$. The bottom plot shows the same pressure in QCD coordinates $(\mu,T)$, featuring critical points located at $(\mu_{BC}=\pm350~\text{MeV},~~T_C=140.1~\text{MeV})$. In all panels, the black dot represents the critical point and the red solid line denotes the first-order transition line.
  } \label{fig:MappingIsing}
\end{figure*}

\begin{figure}[htbp]
    \centering
    \includegraphics[width=1\linewidth]{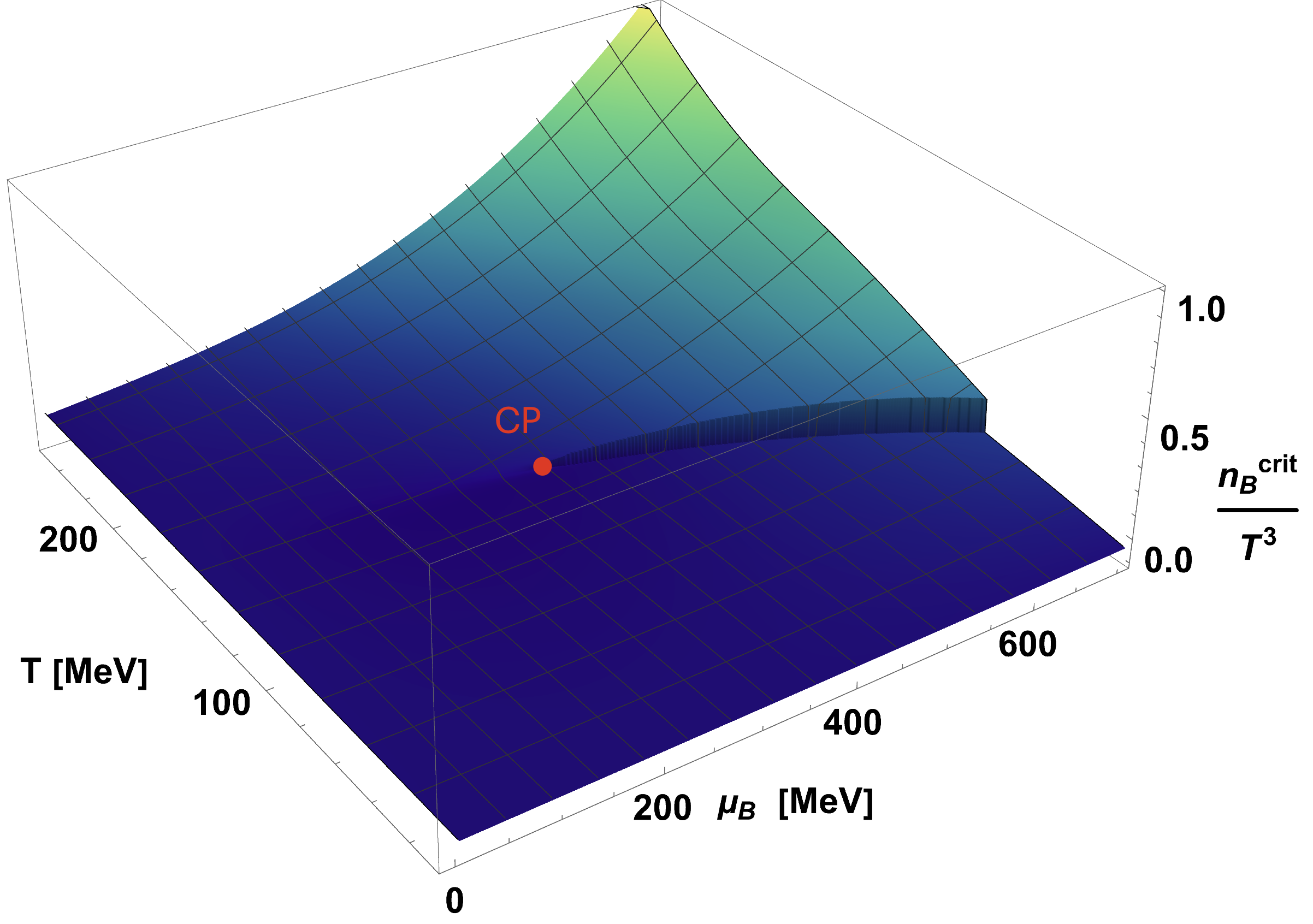}
    \caption{Critical baryon density for the chosen parameters $w = 2, \rho = 2$, $\alpha_{12}=90^0$ with the critical point at $\mu_{BC} = 350$ MeV and $T_C = 140 ~\text{MeV}$. For $\mu_B < \mu_{BC}$, no significant changes occur, indicating a smooth crossover transition. However, for $\mu_B > \mu_{BC}$, a distinct jump appears, marking the transition as first-order. } 
    \label{fig:baryon_crit}
\end{figure}

 In this paper, we illustrate two choices of critical baryon chemical potential. The first one, used mainly for comparison with the BEST collaboration EoS, is $\mu_{BC}=350 ~\text{MeV}$, giving $T_C =140 ~\text{MeV}$ and $\alpha_1 = 6.7^\circ $. For the first choice of parameters, we show in Fig.~\ref{fig:MappingIsing} contours of equal normalized critical pressure, in the $r-h$, $T^\prime-\mu_B^2$ and $T-\mu_B$ planes. The first-order transition line is shown as a red solid line, and the critical point corresponds to a black dot. We show positive and negative values of $\mu_B$, corresponding to positive and negative baryon chemical potentials, to illustrate the symmetry of QCD under baryon-and-antibaryon exchange. This symmetry arises naturally from the selection of a quadratic mapping of the chemical potential in Eq. \eqref{quadrticmappingEq}. For the same choice of parameters, in Fig.~\ref{fig:baryon_crit} we show the critical baryon density, which develops a discontinuity for $\mu_B > \mu_{BC}$, as required for a first order transition.  
 With the second choice we place the critical point in a region that goes beyond the limits of the BEST collaboration EoS: we choose $\mu_{BC}=500~ \text{MeV}$, corresponding to $T_C =116~ \text{MeV}$ and  $\alpha_1 = 11.2^\circ $.  
 
\section{\label{sec:merging}Equation of State: Merging the lattice data and the critical point singularity}

It is important to keep in mind that equation~\eqref{Eq:alternativeExs} is the {\em definition} of $T'(T,\mu_B)$.
Since the function $\chi_2^B(T',0)$ is analytic (smooth crossover), the singularity in $n_B$ due to the critical point and the first-order transition must be carried by $T'(T,\mu_B)$.
Since the singularity of $n_B$ is inherited from the singularity of the pressure via Eq.~\eqref{eq:chi1-crit}, we can determine the corresponding singularity in $T'$ via equation~\eqref{Eq:alternativeExs}. 

We shall separate the baryon density into a regular and singular parts: $n_B=n_B^{\rm reg}+n_B^{\rm crit}$, where $n_B^{\rm crit}$ is defined by Eq.\eqref{eq:chi1-crit}. 
Similarly, we separate $T'$: $T'=T'_{\rm reg}+T'_{\rm crit}$.  
Since $n_B^{\rm crit}$ vanishes at the critical point we can expand $\chi_2^B$ in Eq.\eqref{Eq:alternativeExs} and obtain the relationship between $T'_{\rm crit}$ and $n_B^{\rm crit}$:
\begin{eqnarray}
    T'_{\rm crit}(T,\mu_B) =   \left(\frac{\partial \chi^B_{2}(T,0)}{\partial T}\Big|_{T_0}\right)^{-1} \frac{n^{\rm crit}_B(T,\mu_B)}{T^3 \times(\mu_B/T)} 
    \label{Eq:Tcrit}
\end{eqnarray}
Of course, the Taylor expansion of $T'_{\rm crit}$ is different from Eq.\eqref{Shifting} inferred from lattice data. However, we can always choose the {\em regular} $T'_{\rm reg}$ contribution so that the Taylor expansion of the {\em full} $T'$ agrees with the lattice.
To match lattice results at low 
$(\mu_B/T)$, since $\kappa_4^{BB}(T)$ is consistent with zero, we can truncate the Taylor expansion in Eq.\eqref{Shifting} and define 
\begin{equation}
    T'_{\rm lat}(T,\mu_B) = T\left(1+\kappa_2^{BB}(T)\left(\frac{\mu_B}{T}\right)^2\right).
\end{equation}
We can then write
\begin{align} \nonumber
    T'
    (T,\mu_B) &= \underbrace{T'_{\rm lat}(T,\mu_B)}_{\text{lowest orders in $(\mu_B/T)$ }} + \\ 
    & \qquad \underbrace{T'_{\rm crit}(T,\mu_B) - \text{Taylor}_{n\le2}[T'_{\text{\rm crit}}(T,\mu_B)]}_{\text{higher orders in $(\mu_B/T)$ }} \, \, ,
    \label{Tfull}
\end{align}
which has the same singularity as $T'_{\rm crit}$ and the same {\em truncated} Taylor expansion as $T'_{\rm lat}$.

The last term in Eq.\eqref{Tfull} represents the Taylor-expansion of 
$T'_{\rm crit}(T,\mu_B)$, which we will carry out to order $\mathcal{O}((\mu_B/T)^2)$ and truncate beyond that order. Using Eq.\eqref{Eq:Tcrit} we find:
\begin{align}
    \text{Taylor}_{n\le2}[T'_{\text{\rm crit}}] &= \left(\frac{\partial \chi_{2,\rm lat}^B(T)}{\partial T}\bigg|_{T_0}\right)^{-1}\bigg[\frac{\partial n_B^{\text{\rm crit}}/T^3}{\partial (\mu_B/T)}\bigg|_{\hat{\mu}_B=0} \nonumber \\
    &\quad + \frac{1}{3!}\frac{\partial^3 n_B^{\text{\rm crit}}/T^3}{\partial (\mu_B/T)^3}\bigg|_{\hat{\mu}_B=0}\left(\frac{\mu_B}{T}\right)^2\bigg].
\end{align}
One can thus identify the regular contribution $T'_{\rm reg}$, using Eq.\eqref{Tfull}, as $T'_{\rm reg} = T'_{\rm lat} - \text{Taylor}_{n\le2}[T'_{\text{\rm crit}}]$.

\begin{figure}[!htbp]
    \centering
    \includegraphics[scale=0.18]{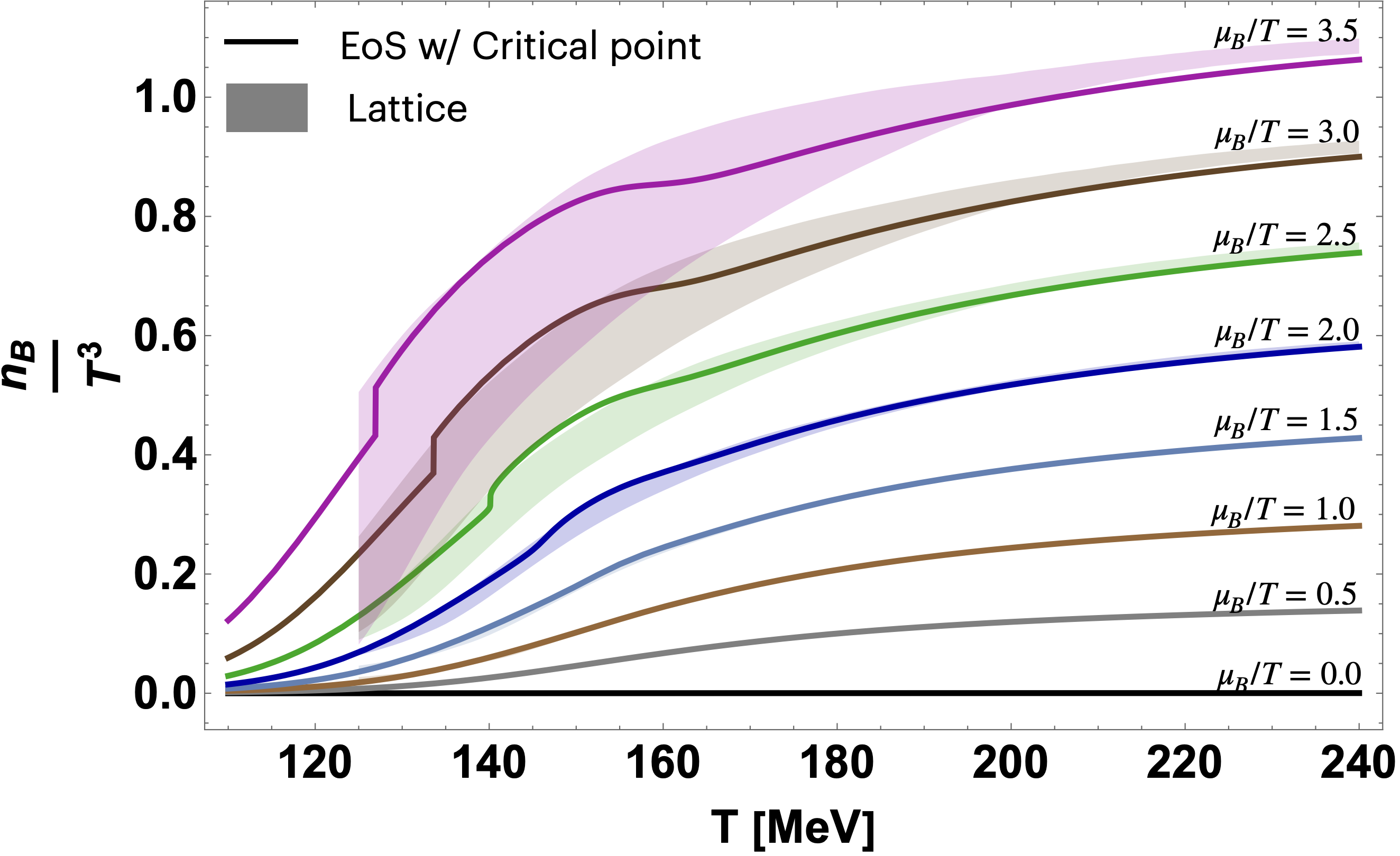}
    \caption{Baryon density as a function of the temperature, for different values of $\mu_B/T$. 
    Solid lines correspond to the equation of state with a critical point located at $\mu_{BC} = 350$ MeV resulting in $T_C =140 ~\text{MeV},~ \alpha_1 = 6.65^0$, with $\alpha_2 = \alpha_1 - \alpha_{12}, ~ \alpha_{12}=90^0, ~ w=2$ and $\rho =2 $. They are compared to lattice QCD results from Ref. \cite{Borsanyi:2021sxv} obtained using the $T'$-expansion scheme, shown as bands indicating the errors due to Taylor expansion truncation.
    }
    \label{fig:lattice_Ising}
\end{figure}

At this point, inserting Eq.~\eqref{Tfull} in \eqref{Eq:alternativeExs} 
completely defines the baryon density with a critical point for a
chosen set of critical point parameters. As an example, we show in Fig.~\ref{fig:lattice_Ising} the baryon 
density as a function of the temperature, for different values of $\mu_B/T$, for a critical point located at 
$\mu_{BC}=350$ MeV, resulting in $T_C=140 $ MeV, $\alpha_1=6.65^\circ$, 
with $\alpha_2=\alpha_1-\alpha_{12}$, $\alpha_{12}=90^\circ$, $w=2$ 
and $\rho=2$. We compare these results with lattice QCD results obtained 
in Ref.~\cite{Borsanyi:2021sxv} from the alternative expansion scheme. 
Notably, we can see that our results are not in tension, within error 
bars, with the lattice ones, even when a critical point is placed in the
chemical potential regime accessible to the extrapolation.

\section{\label{sec:Thermodynamics}Results : Thermodynamics}
In this section, we calculate all thermodynamic observables. From 
Eq.~(\ref{Eq:alternativeExs}), the baryon density $n_B(T,\mu_B)$ in temperature 
and chemical potential is readily provided, and the pressure $P(T,\mu_B)$ is obtained through simple integration:
\begin{equation}
    \frac{P(T,\mu_B)}{T^4} = \chi_{0,\rm lat}^B(T,0) + \frac{1}{T}\int_0^{\mu_B}d\mu_{B}^\prime\frac{n_B(T,\mu_{B}^\prime)}{T^3} \, \, .
\end{equation}
The integration constant $\chi_{0,\rm lat}^B(T,0)$ is the pressure
at $\mu_B=0$, for which we employ lattice QCD results from Ref.~\cite{Borsanyi:2013bia}.

\begin{figure}[!htbp]
    \centering
    \includegraphics[scale=0.2]{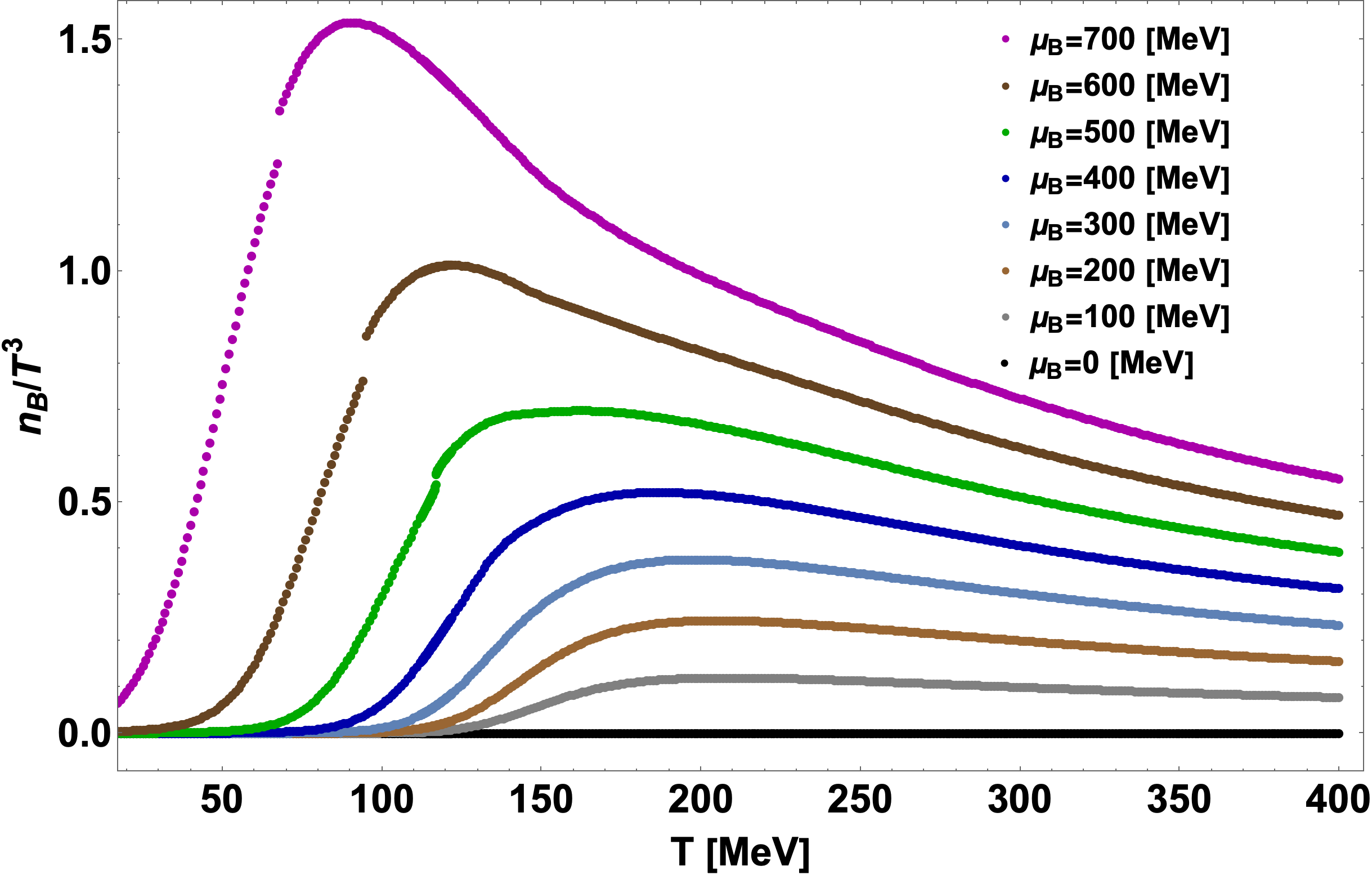}
    \caption{Baryon density as a function of the temperature for different baryon chemical potentials. As expected, a discontinuity appears when $\mu_B > \mu_{BC}$, where the transition is first order. The critical point is located at $\mu_{BC}=500 ~\text{MeV}$, resulting in $T_C=117~ \text{MeV}$ and $\alpha_1 = 11^\circ$. Additionally, we have $w =15$, $\rho=0.3$, and $\alpha_{12} =\alpha_1 $, meaning $\alpha_{2} = \alpha_1 - \alpha_{12}=0$. }
    \label{fig:Baryondensity2D}
\end{figure}

\begin{figure}[!htbp]
    \centering
    \includegraphics[scale=0.26]{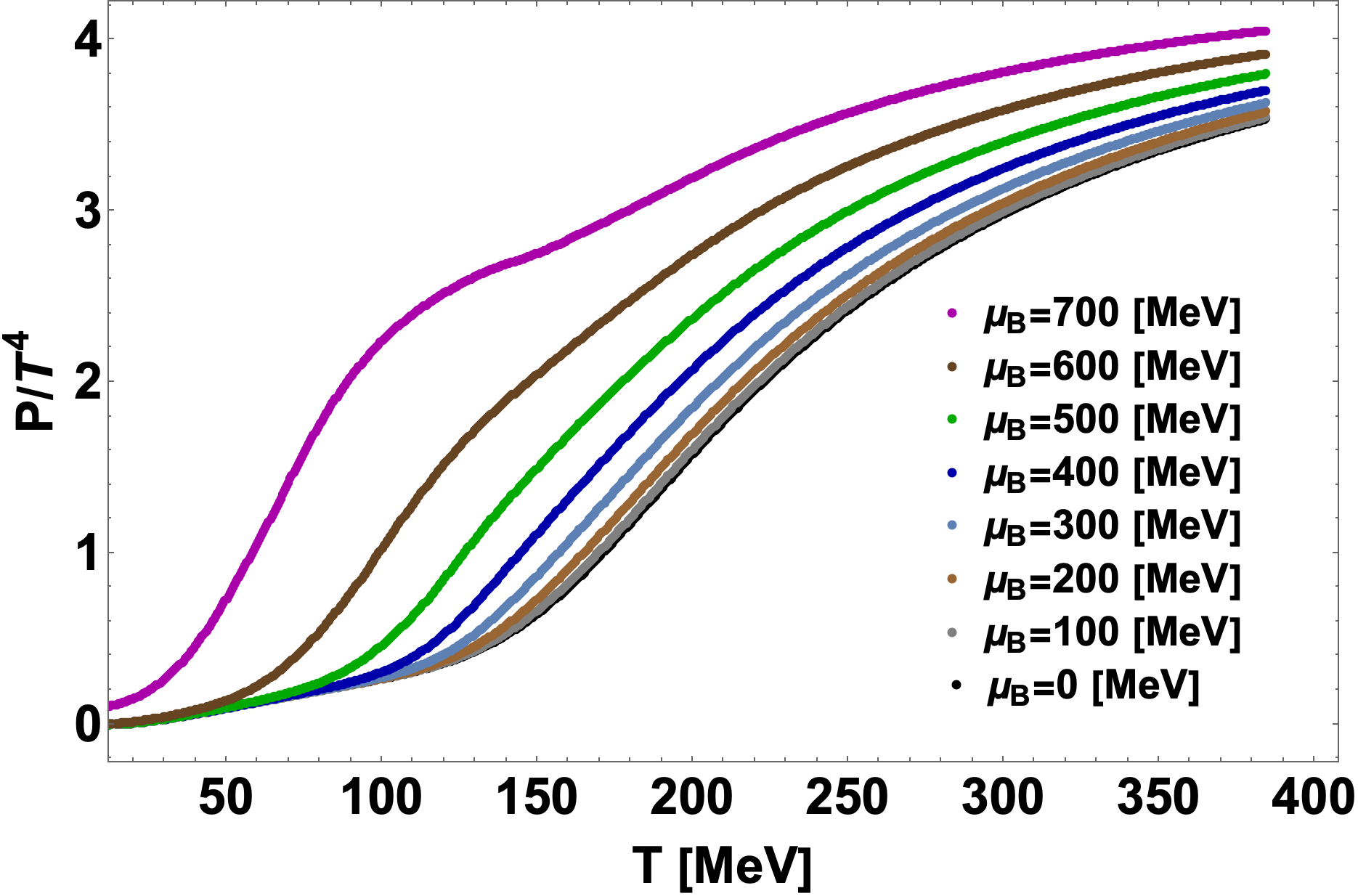}
    \caption{Pressure as a function of the temperature for different baryon chemical potentials. The critical point manifests itself less clearly in the pressure, which only develops a kink for $\mu_B>\mu_{BC}$.
    The plot corresponds to the same parameters as the ones used in Fig.~\ref{fig:Baryondensity2D}.}
    \label{fig:Pressure2D}
\end{figure}

\begin{figure}[!htbp]
    \centering
    \includegraphics[scale=0.27]{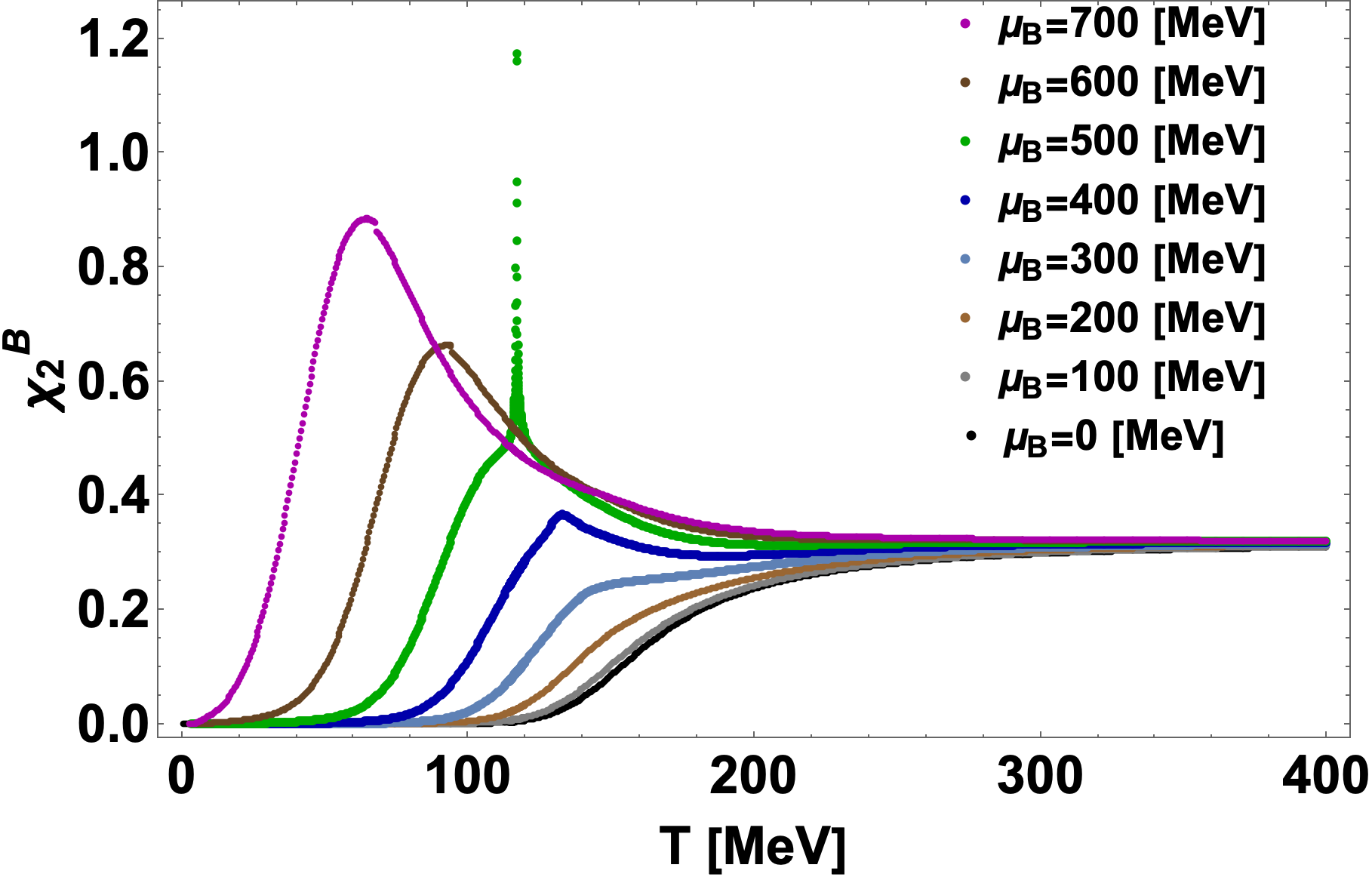}
    \caption{Second order baryon susceptibility as a function of the temperature for different baryon chemical potentials. This quantity represents the measure of how baryon density reacts to an increase in chemical potential. A divergence is expected at the critical point, which can be seen for $\mu_B=\mu_{BC}= 500 \, \text{MeV}$. The plot corresponds to the same parameters as the ones used in Fig.~\ref{fig:Baryondensity2D}. }
    \label{fig:Chi2density2D}
\end{figure}

\begin{figure}[!htbp]
    \centering
    \includegraphics[scale=0.28]{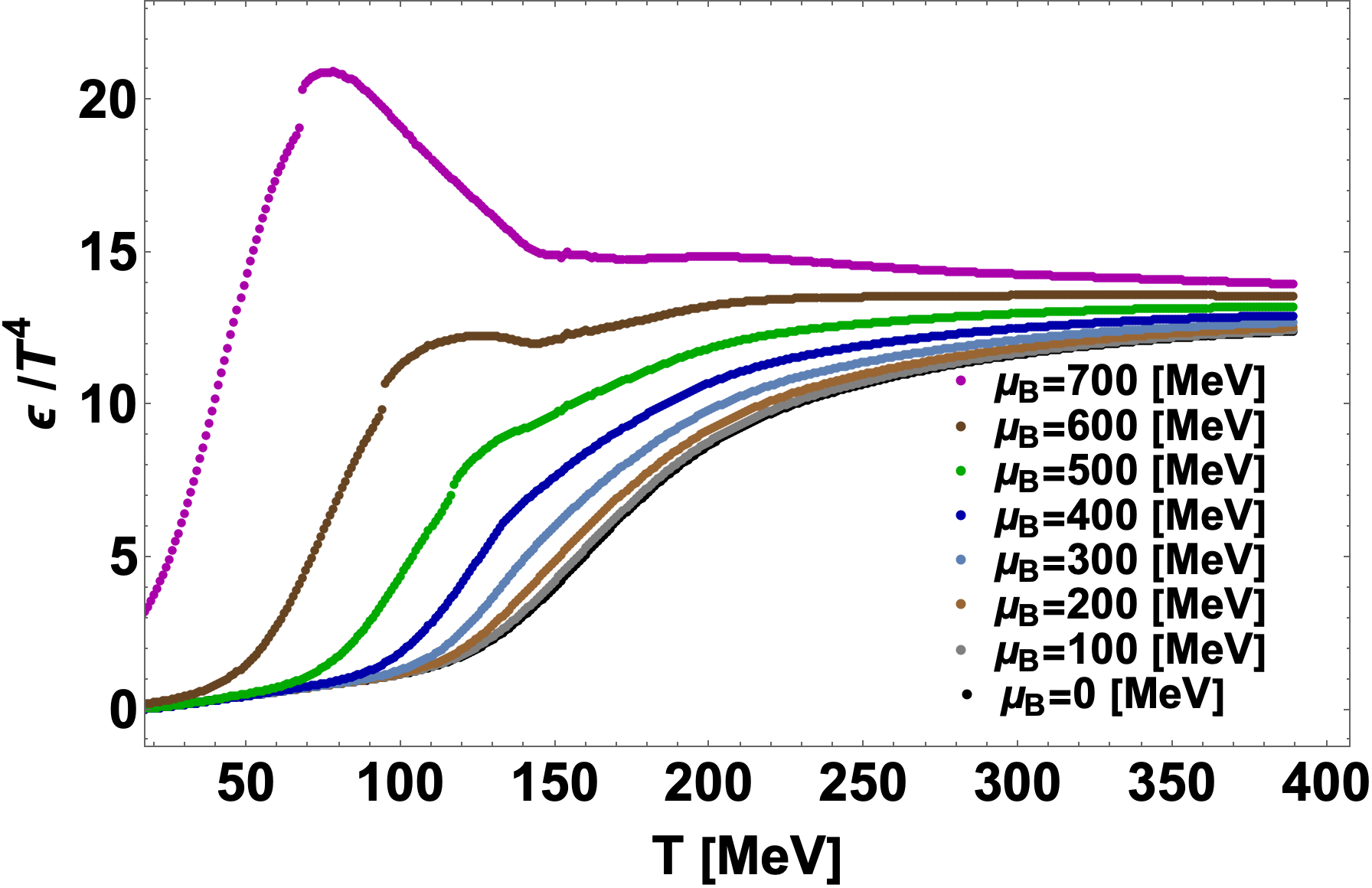}
    \caption{Energy density as a function of the temperature for different baryon chemical potentials. This quantity also shows a discontinuity for $\mu_B>\mu_{BC}$. The plot corresponds to the same parameters as the ones used in Fig.~\ref{fig:Baryondensity2D}.}
    \label{fig:Energydensity3D}
\end{figure}

Entropy density, energy density and second baryon susceptibility are derivatives of pressure and baryon density, defined as:
\begin{eqnarray}
    \frac{s(T,\mu_B)}{T^3} &&= \frac{1}{T^3}\frac{\partial P(T,\mu_B)}{\partial T}\Big|_{\mu_B}\\
   \frac{\epsilon(T,\mu_B)}{T^4} &&= - \frac{P(T,\hat{\mu}_B)}{T^4} + \frac{s(T,\hat{\mu}_B)}{T^3}  + \hat{\mu}_B\frac{n_B(T,\hat{\mu}_B)}{T^3}  \\
    \chi_2^B(T,\mu_B) &&= \frac{\partial ({n_B(T,\mu_B)}/T^3)}{\partial \mu_B/T}\bigg|_{T}
\end{eqnarray}
which we implement through Eqs.~\eqref{Eq:dnBdmuB} and~\eqref{Eq:dnBdT}. In Figs.~\ref{fig:Baryondensity2D},~\ref{fig:Pressure2D},~\ref{fig:Chi2density2D}, and~\ref{fig:Energydensity3D} we show the baryon density, pressure, second baryon susceptibility and energy density, respectively, as functions of the temperature, for different values of the baryon chemical potential. These correspond to a critical point located at $\mu_{BC}=500 ~\text{MeV}$, resulting in $T_C=117~ \text{MeV}$ and $\alpha_1 = 11^\circ$. Additionally, we have $w =15$, $\rho=0.3$, and $\alpha_{12} =\alpha_1 $, meaning $\alpha_{2} = \alpha_1 - \alpha_{12}=0$.

\section{\label{sec:Constraint} Constraints on the EoS}

In this manuscript, we obtain a family of equations of state which depend on the free parameters $\mu_{BC}$, $w$, $\rho $ and $\alpha_{12}$ introduced by the mapping in Eq. (\ref{fig:Mapping}). However, the values of these parameters can be guided by physics and the current knowledge from experiments, in order to constrain them and obtain a physical equation of state that describes strongly interacting matter.

\subsection{\label{sec:results}Lattice Results}
While our equations of state depend on the free parameters at high $\mu_B$, we require that they all reproduce lattice QCD results for pressure and its $\mu_B$ derivatives up to 4th order at $\mu_B=0$. 
This can be inferred from Fig. \ref{fig:lattice_Ising}, where we compare our baryon density (exhibiting a discontinuity at $\mu_B>\mu_{BC}$) with the lattice QCD results from Ref. \cite{Borsanyi:2021sxv}: within error-bars, our discontinuity does not contradict the results from lattice QCD.

\subsection{\label{sec:masses}Physical Quark masses}
In Ref. \cite{Pradeep:2019ccv}, a thorough investigation was conducted regarding the linear mapping from Ising to QCD introduced in Refs. \cite{Parotto:2018pwx,Karthein:2021dcn}. This study effectively explored the scenario in which the critical point closely approaches the tricritical point, revealing a universal dependence of the mapping parameters on the quark mass $m_q$. Notably, when the critical point resides in the proximity of the tricritical point, the angle denoted as $\alpha_{12}$ between the lines of $r=0$ and $h=0$ within the $(T,\mu_B)$ plane decreases, exhibiting a behavior proportional to $m_q^{2/5}$. For a physical quark mass $m_q$, the angle $\alpha_{12} \approx \alpha'_{12}  $ as in \eqref{alpha12p_alpha12}  needs to be small, approximately equal to  $  \alpha_1$.

\subsection{\label{sec:stability}Stability and Causality }
The non-universal mapping from $(r, h)$ to $(T, \mu_B)$ leaves open the selection of free parameters. While the angle $\alpha_{12}$ can be constrained by the physical value of the quark masses, there is no physical guidance for the scaling parameters $(w, \rho)$. Potentially, some choices of parameters would lead to an unstable equation of state.

\begin{figure}[!htbp]
    \centering
    \includegraphics[scale=0.28]{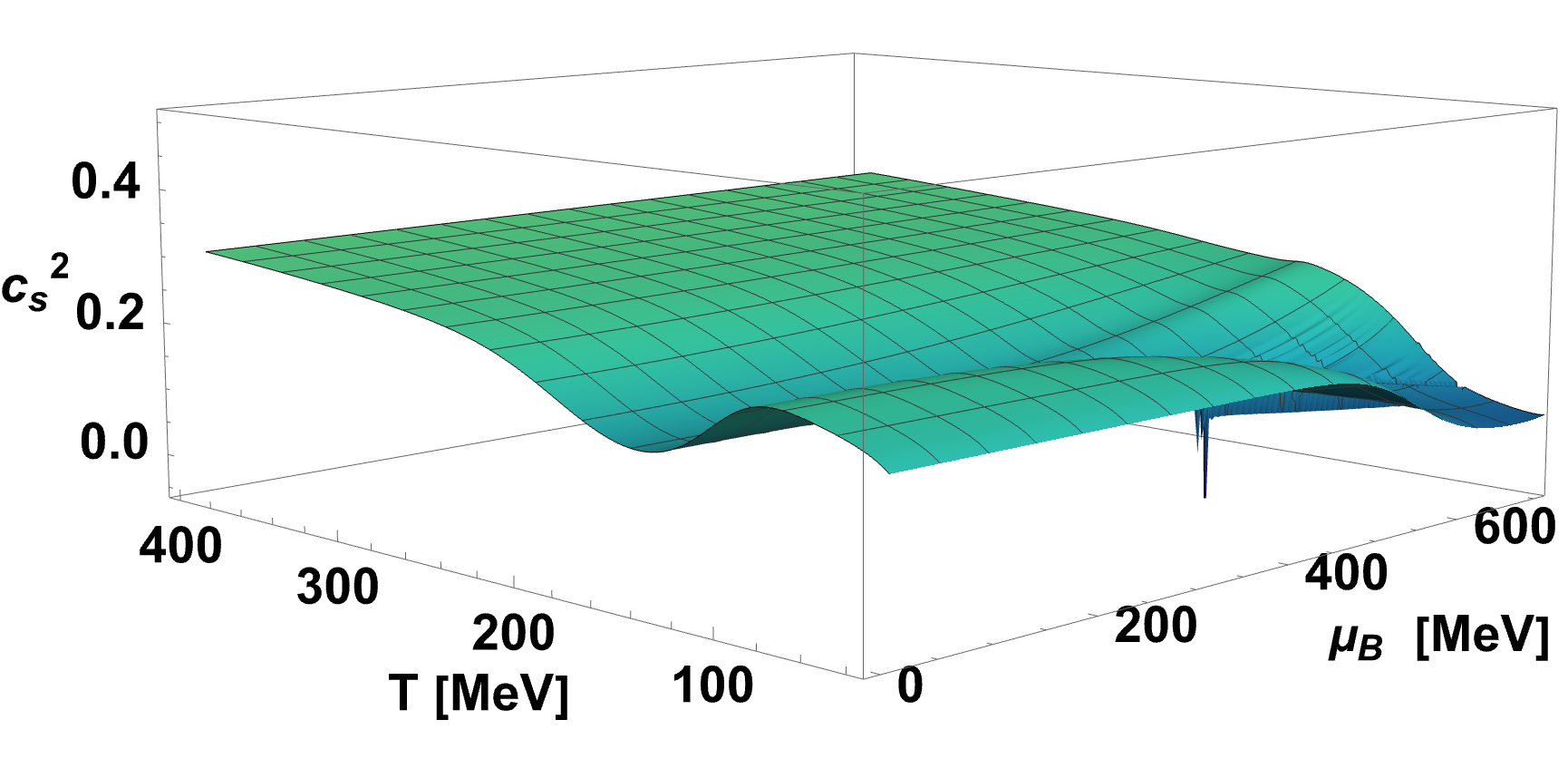}
    \caption{ Speed of sound as a function of temperature and baryon chemical potential. This quantity shows a pronounced at the critical point.
    The plot corresponds to the same parameters as the ones used in Fig: \ref{fig:Baryondensity2D}.\label{fig:Speedofsound}
    }
    \label{fig:Sp_Sound3D}
\end{figure}

\begin{figure}[!htbp]
    \centering
    \includegraphics[scale=0.22]{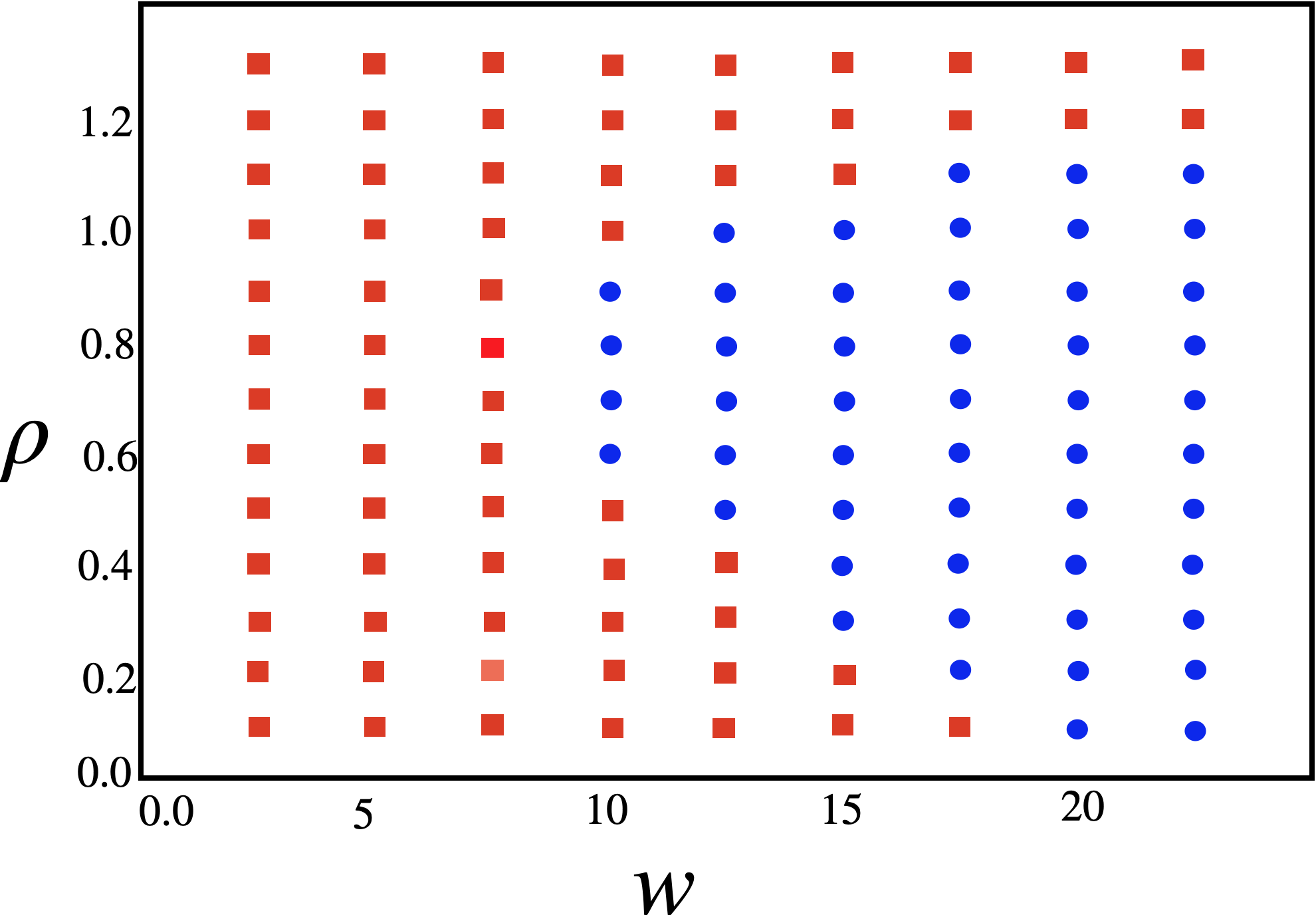}
    \caption{Landscape plot in the range $w = 2.5 - 22.5$, $\rho = 0.1 - 1.3$ at fixed values of $\mu_{BC}=500 ~\text{MeV}, ~T_C = 117 ~\text{MeV},~\alpha_1 = 11^0$ and $\alpha_{12} = \alpha_1$. Red squares correspond to a pathological choice of parameters in the range $\mu_B = \{0, 700~ \text{MeV}\}$, while the blue dots represent acceptable ones.
}
    \label{fig:Tmappingparameter5}
\end{figure}

For a valid equation of state, certain conditions must be met. We require that the pressure is a monotonically increasing function of $T$ and $\mu_B$, which means positivity of baryon density, entropy density, energy density, speed of sound, and baryon number susceptibility everywhere in the $(T,\mu_B)$ plane \cite{Mroczek:2022oga} ranging from $0<\mu_B<700$ MeV and $25 \, \text{MeV} < T< 800$ MeV. This can be summarized in two conditions: positivity of the second baryon susceptibility $\chi_2$ and of the specific heat at constant volume $c_V$, which can be written as \cite{Floerchinger:2015efa}:
\begin{eqnarray}
    c_V(T,\mu_B) = &\frac{T}{\chi_2^B} \left[ \frac{\partial s}{\partial T}\chi_2^B- \left(\frac{\partial n_B}{\partial T}\right)^2\right] \, \, . \label{cv}
\end{eqnarray}
Additionally, to uphold causality, the speed of sound 
\begin{eqnarray}
    c_s^2(T,\mu_B) = \left(\frac{\partial p}{\partial \epsilon}\right)_{s/n} = \frac{n^2\frac{\partial^2p}{\partial T^2} - 2sn\frac{\partial^2p}{\partial T\partial \mu_B} + s^2\frac{\partial^2p}{\partial \mu_B^2}}{(\epsilon + p)\left(\frac{\partial^2p}{\partial T^2}\frac{\partial^2 p}{\partial \mu_B^2} -\left(\frac{\partial^2 p}{\partial T\partial \mu_B}\right)^2\right)} \, \, 
    \nonumber\\
\end{eqnarray}
must fall within the range $0\leq c_s^2\leq 1$. The behavior of the speed of sound as a function of $T$ and $\mu_B$ can be seen in Fig. \ref{fig:Speedofsound}. It exhibits a dip at the critical point, where it vanishes. We show in Fig.~\ref{fig:Tmappingparameter5} a landscape of acceptable (blue dots) and pathological (red squares) choices for the parameters $w$ and $\rho$, for a critical point located at $\mu_{BC}=500 ~\text{MeV}$, which corresponds to $T_C=117~ \text{MeV}$ and $\alpha_1 = 11^\circ$. Additionally, we have $\alpha_{12} =\alpha_1 $, meaning $\alpha_{2} = \alpha_1 - \alpha_{12}$, while $w$ and $\rho$ are varied in the range $w = 2.5 - 22.5$, $\rho = 0.1 - 1.3$. 
Similar plots, comparing our parameter landscapes to the ones from the BEST collaboration EoS, are discussed in Appendix \ref{appendixC}.

\section{\label{sec:conclusion}Summary and Conclusions}

Determining the QCD equation of state, in particular, establishing the existence of the QCD critical point and pinning down its location, is a major goal of heavy-ion collision experiments. The strategy based on comparing predictions of hydrodynamics sensitive to EoS  with experiment requires a parametric family of EoS which can be fed into a hydrodynamic code. In this paper, we introduced a novel framework for constructing such a family of
QCD equations of state. 

Our framework improves on the BEST collaboration approach \cite{Parotto:2018pwx} by introducing several significant innovations. This allows us to achieve coverage over a wider range of the QCD phase diagram relevant for critical point searches.

The main innovation in our paper is merging the universal critical point singularity with the implementation of the $T'$-expansion scheme \cite{Borsanyi:2021sxv}. The $T'$-expansion scheme takes into account the observation that the temperature driven crossover looks remarkably similar at different chemical potentials, the main difference being a shift of the crossover temperature with increasing $\mu_B$. The ``rescaled temperature'' $T'(T,\mu_B)$ defined in Eq.\eqref{Eq:alternativeExs} carries information about the dependence of the position and the shape of the crossover at different $\mu_B$. Since this dependence is relatively slow, the expansion of $T'$, as in Eq.\eqref{Shifting}, is much better controlled than the expansion of quantities such as $\chi_2^B$, which vary rapidly at the crossover.

We introduce the critical singularity into the function $T'(T,\mu_B)$, while making sure that the Taylor expansion coefficients (at $\mu_B=0$) still agree with the lattice data.

Another innovation, relative to the BEST EoS framework, is the mapping of the Ising coordinates $r$ and $h$ into QCD coordinates $T$ and $\mu_B^2$, instead of $\mu_B$. This takes care of the charge conjugation symmetry and the associated curvature of the QCD pseudocritical line.

We check the novel framework by calculating quantities which must obey thermodynamic inequalities. Of course, for sufficiently large $\mu_B$ or for sufficiently strong critical point singularity, the framework will show its limitations by violating these inequalities. However, the range of parameters where the novel framework is thermodynamically consistent is larger than the same range for the BEST collaboration EoS family. 

In particular, our framework allows us to provide thermodynamically consistent EoS in the range $\mu_B=0-700$ MeV, extending beyond the BEST EoS range $\mu_B=0-450$ MeV. In addition,  the range for critical point parameters $w$ and $\rho$ is also extended compared to the one for the BEST EoS at similar values of $T$ and $\mu_B$.

There are several potential avenues for further improvement. Since the approach is still based on the Taylor expansion, necessarily truncated based on the availability of the lattice data, it inevitably breaks down at sufficiently large $\mu_B$. It might be possible to introduce additional resummation techniques dealing with these limitations at larger $\mu_B$. In addition, the true EoS of QCD possesses the well known periodicity in the complex plane: $\mu_B\to\mu_B+2\pi Ti$ due to the quantization of the baryon number. This periodicity could also be implemented. We leave these and further improvements to future work.

\begin{acknowledgments}
This material is based upon work supported by the National Science Foundation under grants No. PHY-2208724, PHY-1654219 and PHY-2116686, within the framework of the MUSES collaboration, under grant
number No. OAC-2103680 and by the
National Aeronautics and Space Agency (NASA)  under Award Number 80NSSC24K0767. This material is also based upon work supported by the U.S. Department of Energy, Office of Science, Office of Nuclear Physics, under Awards
Number DE-SC0022023, DE-FG02-05ER41367 and DE-FG0201ER41195.
O.S. and E.B. acknowledge support by the Deutsche Forschungsgemeinschaft (DFG, German Research Foundation) through the grant CRC-TR 211 'Strong-interaction matter under extreme conditions' - Project number 315477589 - TRR 211. 
This work is supported by the European Union’s Horizon 2020 research and innovation program under grant agreement No 824093 (STRONG-2020).
\end{acknowledgments}

\appendix

\section{Taylor expansion of the critical contribution}
\label{appendixA}
Here we present the formula and derivation for
$\text{Taylor}[T'_{\rm crit}(T,\mu_B)]$.
Since in our approach $T'_{\rm lat}(T,\mu_B)$ is truncated  up to $\kappa_2^{BB}(T)$, we need the $
\text{Taylor}[T'_{crit}(T,\mu_B),n=2]= a_0(T) + a_2(T)\left(\frac{\mu_B}{T}\right)^2$,
such that we match that same order by construction, while the higher order contributions come from the critical part.   The coefficients $a_0$ and $a_2$ are then given by;
\begin{align}
    a_0(T) &=\left(\frac{\partial \chi_2^B}{\partial T}\bigg|_{T_0}\right)\left(\frac{\partial n_B^{\rm crit}(T,\mu_B)}{\partial (\mu_B/T)}\right)\bigg|_{\mu_B/T=0}\\
    a_2(T) &=\left(\frac{\partial \chi_2^B}{\partial T}\bigg|_{T_0}\right)\left(\frac{1}{3!}\frac{\partial^3 n_B^{\rm crit}(T,\mu_B)}{\partial (\mu_B/T)^3}\right)\bigg|_{\mu_B/T=0}
\end{align}

\section{Computing thermodynamics}
\label{appendixB}
\begin{figure*}[!htbp] 
    \centering
    \includegraphics[width=1.0\textwidth]{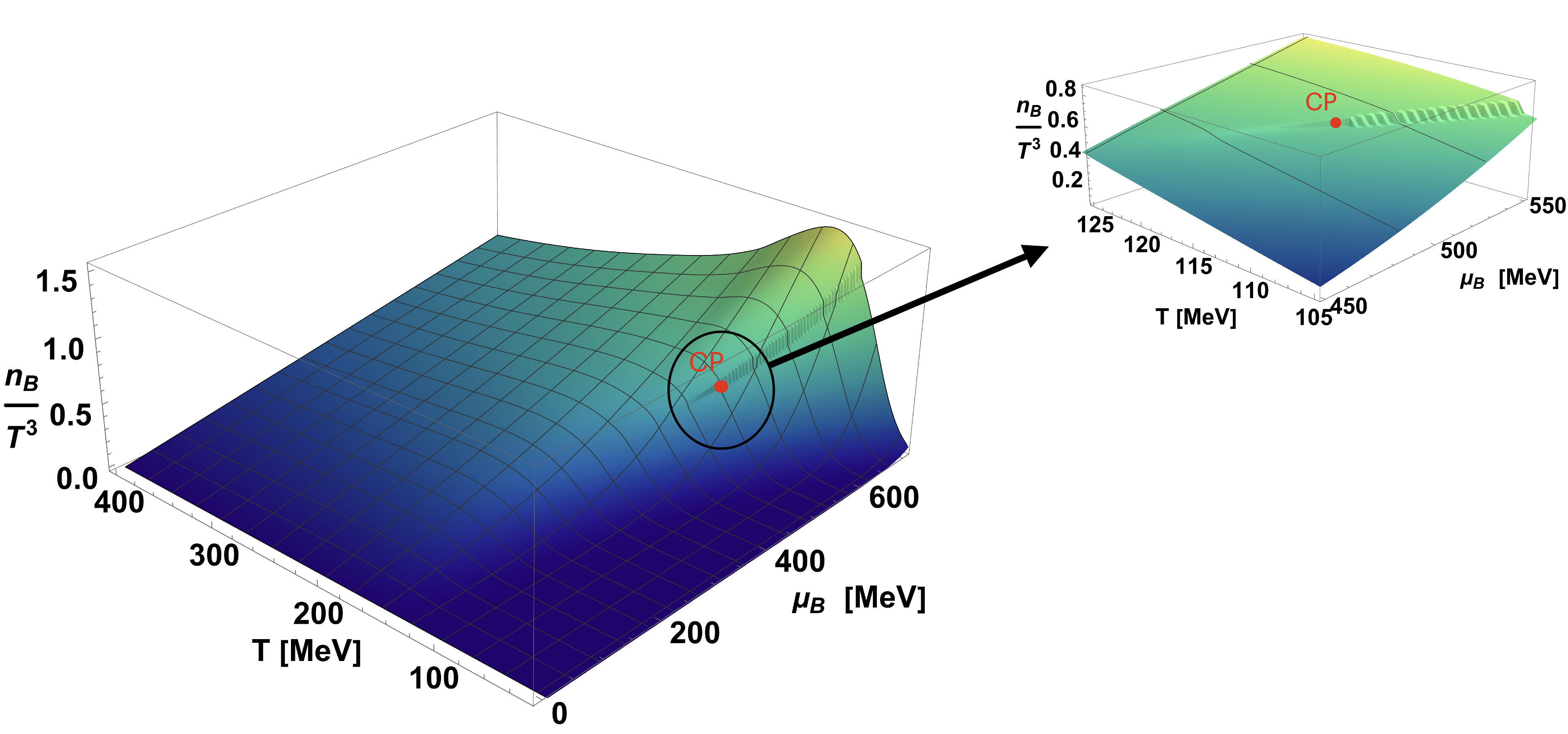}
    \caption{Baryon density as a function of temperature and chemical potential for the same parameters as in Fig.~\ref{fig:Baryondensity2D}, with a zoom into the critical region.}
    \label{fig:Baryondensity3D}
\end{figure*}

From Eq. \eqref{Eq:alternativeExs}, we obtain Eq. (\ref{Eq:dnBdmuB}) and Eq. (\ref{Eq:dnBdT}), which are derivatives of the baryon density with respect to chemical potential and temperature, respectively
\begin{align}\nonumber
    \frac{\partial n_B(T,\mu_B)}{\partial (\mu_B/T)}=\chi_{2,\rm lat}^B(T') T^3 + \\
    \frac{\mu_B}{T}\frac{\partial \chi_{2,\rm lat}^B(T)}{\partial T}\bigg|_{T'}\frac{\partial T'}{\partial (\mu_B/T)} T^3 
    \label{Eq:dnBdmuB}
\end{align}

\begin{align}\nonumber
    \frac{\partial n_B(T,\mu_B)}{\partial T}= 2\frac{n_B(T,\mu_B)}{T} +\\
    \frac{\mu_B}{T}\frac{\partial \chi_{2,\rm lat}^B(T)}{\partial T} \frac{\partial T'}{\partial T} T^3 \label{Eq:dnBdT}
\end{align}
Then entropy is computed from the integral of Eq. \eqref{Eq:dnBdT} using 
\begin{align}\nonumber
 s(T,\mu_B)&= 4T^3\chi_{0,\rm lat}^B(T) + T^4\frac{\chi_{0,\rm lat}^B(T)}{dT}\\ &\int_0^{{\mu}_B}d{\mu}_B' \frac{\partial n_B(T,\mu_B')}{\partial T}
\end{align} 
All thermodynamic quantities, calculated in this paper as functions of temperature and chemical potential, are shown in Figures \ref{fig:Baryondensity2D}-\ref{fig:Energydensity3D} and \ref{fig:Baryondensity3D}-\ref{fig:Entropydensity3D} for slices at constant $\mu_B$ in the main text and 3D plots in the Appendix, respectively.

\section{Comparison with BEST EoS}
\label{appendixC}
In \cite{Parotto:2018pwx,Pradeep:2019ccv,Mroczek:2022oga}, a linear map from Ising to QCD with six parameters was utilized.

\begin{eqnarray}\nonumber
    T-T_C & = & T_C w(r \rho \sin\alpha_1 + h \sin\alpha_2)\\
    \mu_B - \mu_{BC} & = & T_C w(-r \rho \cos\alpha_1 - h \cos\alpha_2).
    \label{linearmapping}
\end{eqnarray}

By making use of the following equations for the slopes at the critical point:
\begin{align}
    \frac{dT}{d\mu_B}\Big|_{h=0} & = -\tan\alpha_1\\
    \frac{dT}{d\mu_B}\Big|_{r=0} & = -\tan\alpha_2
\end{align}
and linearizing Eq. (\ref{quadrticmappingEq}) around the critical point and using \(T'_{\text{lat}} = T\left(1+\kappa_2^{BB}(T)\left(\frac{\mu_B}{T}\right)^2\right)\),

\begin{align}
    \frac{1}{T'_{,T}}\frac{\Delta T'}{\Delta \mu_B} = \frac{\Delta T}{\Delta \mu_{B}} + \frac{2 \kappa_2^{BB}(T)\mu_B}{T'_{,T} T}
\end{align}

At \(h=0\), we get Eq.\eqref{Eq:alpha1}:

\begin{align}
    \tan\alpha_1 = \frac{2\kappa_2^{BB}(T_C)\mu_{BC}}{T'_{,T} T_C}
\end{align}

At \(r=0\), we get Eq. \eqref{Eq:tan12}:
\begin{align}
    \tan\alpha'_{12} = \tan\alpha_1 - \tan\alpha_2
\end{align}
Using simple trigonometric relations and Eq.\eqref{Eq:wpw} and Eq.\eqref{Eq:rhoprho}, we find \eqref{Eq:w} and \eqref{Eq:rho}.

Then, we approximate for either small angles or large angles:

\begin{itemize}
    \item For small angles:
\begin{subequations}
\begin{align}
&\alpha_{12}'\approx\alpha_{12}\,; \label{alpha12p_alpha12}\\
&w'\approx w\,;\\
&\rho'\approx\rho\,;  
\end{align}    
\end{subequations}

\item For $\alpha_1\ll1$ and $\alpha_{12}=90^\circ$:
   \begin{subequations}
\begin{align}
\alpha_{12}'&\approx 90^\circ-\alpha_1\,;\\
w'&\approx w\,;\\
\rho'&\approx\rho\,.
\end{align}    
\end{subequations}
\end{itemize}

\begin{figure*}[!htbp]
    \centering
    \includegraphics[width=0.47\textwidth]{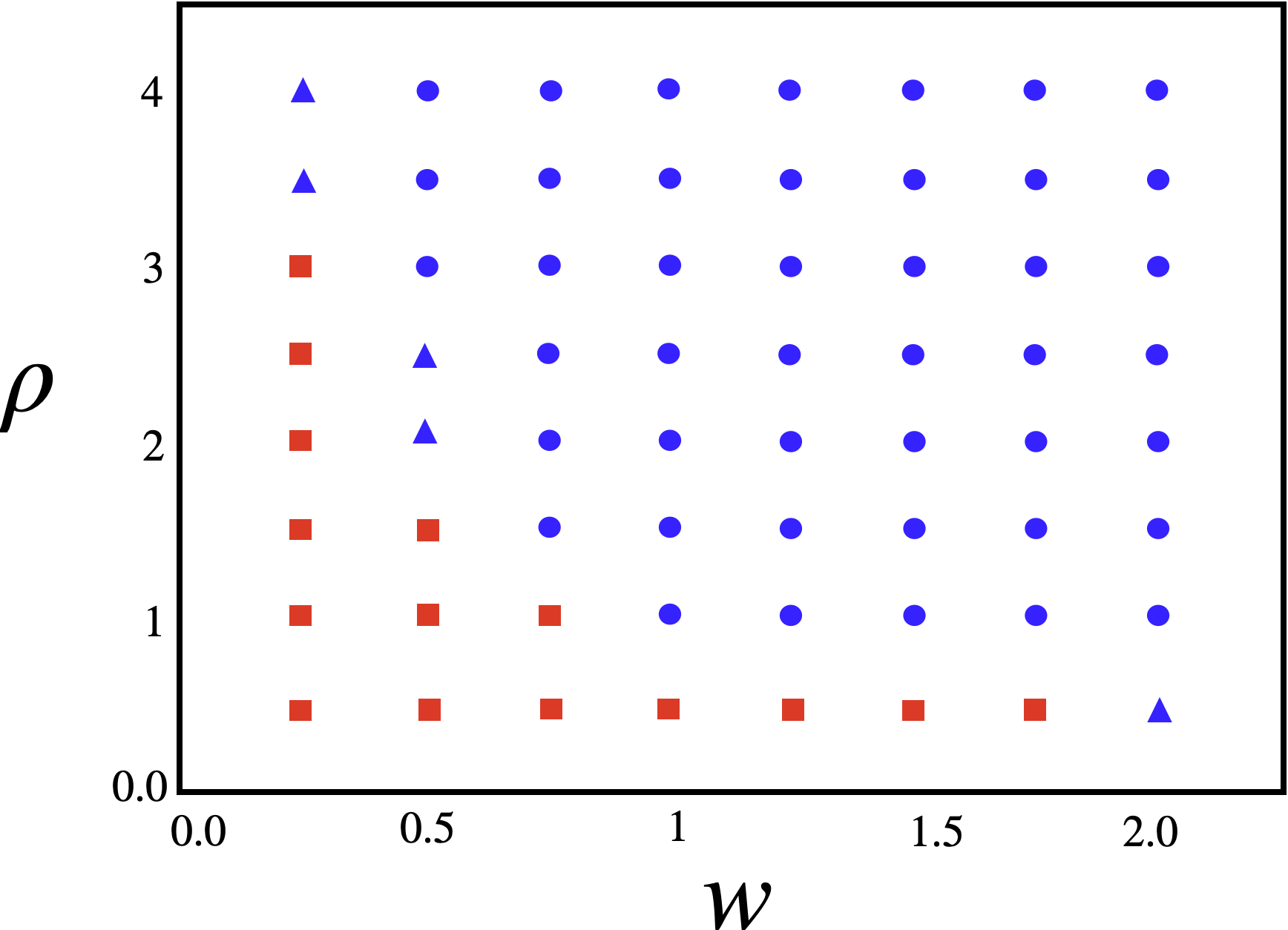}\hfill
    \includegraphics[width=0.45\textwidth]{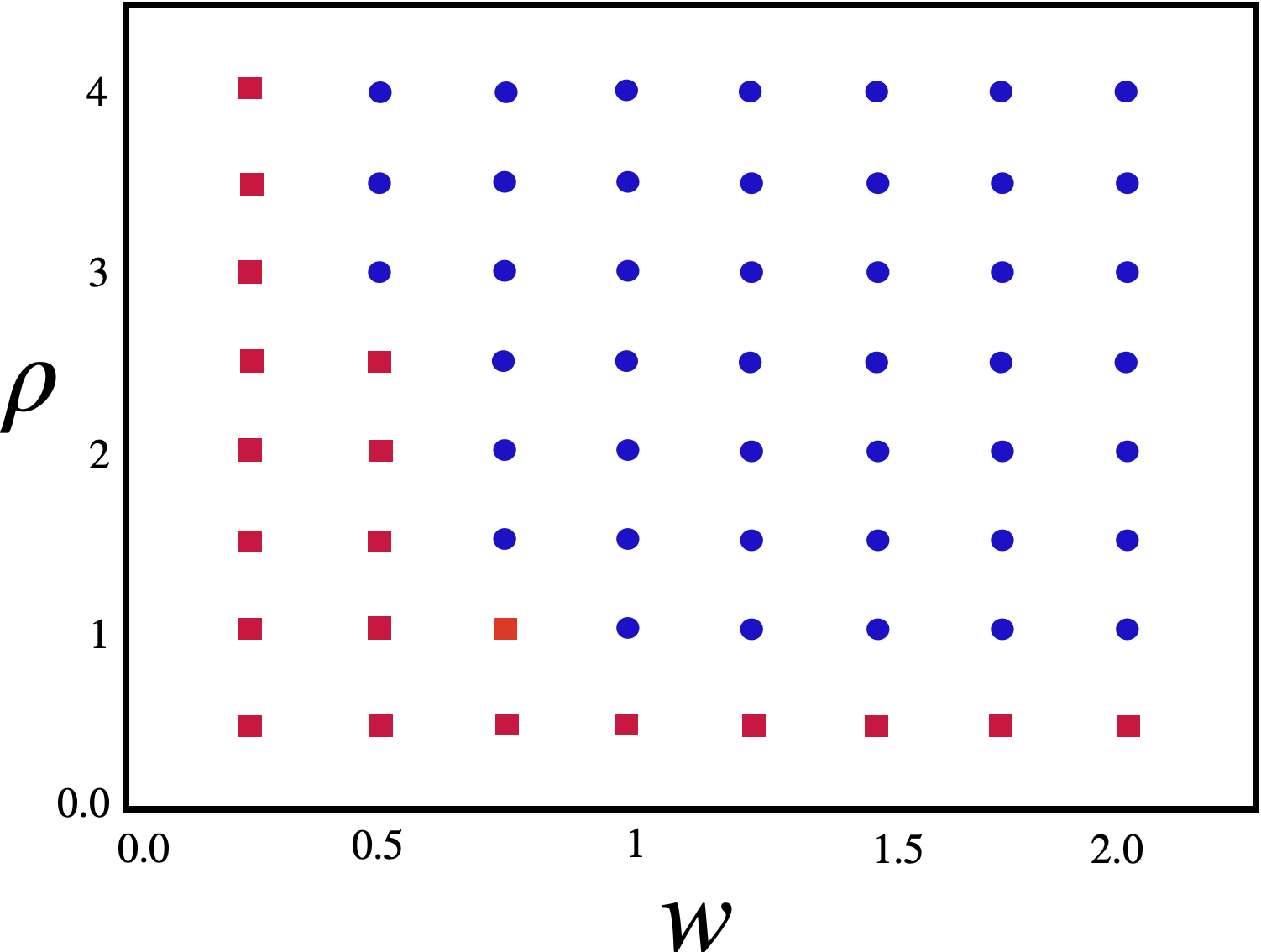}
    \caption{Comparison of the stability plots for \(w\) and \(\rho\) with the new mapping (Quadratic) on the left panel and the BEST mapping (Linear) \cite{Parotto:2018pwx} on the right. The blue points represent acceptable parameters, while the red points denote unacceptable ones for \(\mu_{BC} = 350 \, \text{MeV}, T_C = 140.073 \, \text{MeV}, \alpha_1 = 6.653^\circ, \kappa = 0.02633\) with 
\(    \alpha_{12} = 90^\circ\) for $\mu_B=\{0,450 \text{MeV}\}$.
    Blue triangles on the left panel indicate parameters that are acceptable in the new mapping, while they were not in the BEST collaboration mapping.
    }
    \label{fig:mappingsparamter90}
\end{figure*}

\begin{figure*}[!htbp]
    \centering
    \includegraphics[width=0.48\textwidth]{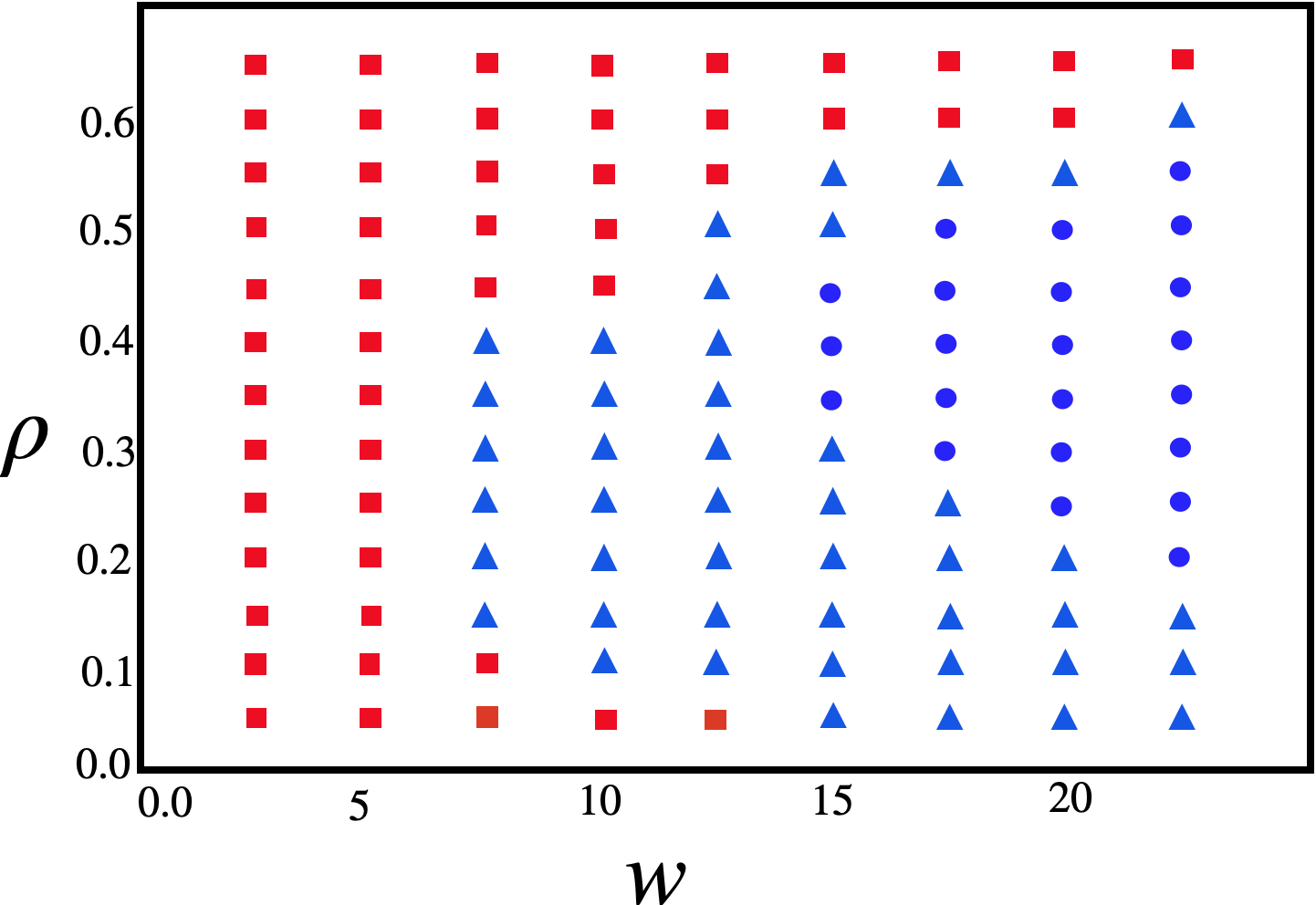}
    \hfill
    \includegraphics[width=0.48\textwidth]{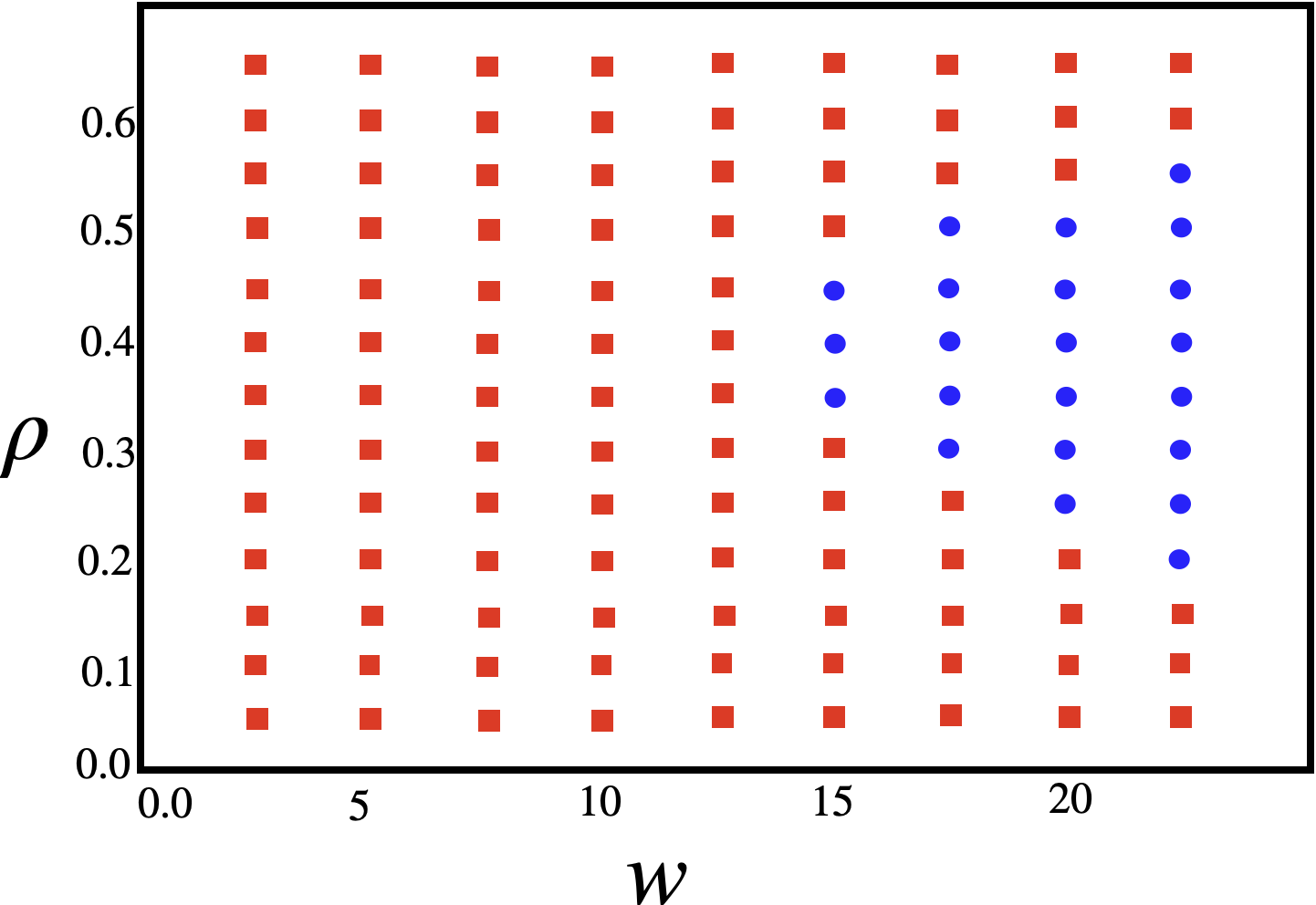}
    \caption{Comparison of the stability plots for \(w\) and \(\rho\) with the new mapping (Quadratic) on the left panel and the BEST collaboration mapping (Linear) \cite{Parotto:2018pwx} on the right. The blue points represent acceptable parameters, while the red points denote unacceptable ones for \(\mu_{BC} = 350 \, \text{MeV}, T_C = 140.073 \, \text{MeV}, \alpha_1 = 6.653^\circ, \kappa = 0.02633\) with \(\alpha_2=0\), i.e.,
    \(\alpha_{12}' = 
    \alpha_{12} = \alpha_1\) for $\mu_B=\{0,450 \text{MeV}\}$.
    Blue triangles on the left panel indicate parameters that are acceptable in the new mapping, while they were not in the BEST collaboration mapping. 
    }
    \label{fig:Parametercompare5}
\end{figure*}

In Figures \ref{fig:mappingsparamter90} and \ref{fig:Parametercompare5}, we compare the stability parameter landscape in $w$ and $\rho$ for our approach to the ones from the BEST collaboration EoS. In Fig. \ref{fig:mappingsparamter90}, the Ising model axes are chosen to be orthogonal to each other.
However, since physically motivated values of the angle $\alpha_{12}$, characterizing the shape of the critical region, are small \cite{Pradeep:2019ccv}, it is important that  the improvement in the $w$ and $\rho$ ranges is especially pronounced for small angle $\alpha_{12}$, as shown in Fig.\ref{fig:Parametercompare5} for $\alpha_2=0$, i.e., $\alpha_{12}=\alpha_1$.
From these figures it is clear that the quadratic mapping in Fig: \ref{fig:Mapping} has more acceptable points than the linear mapping in \cite{Parotto:2018pwx}.

\begin{figure*}[!htbp] 
    \includegraphics[width=1.0\textwidth]{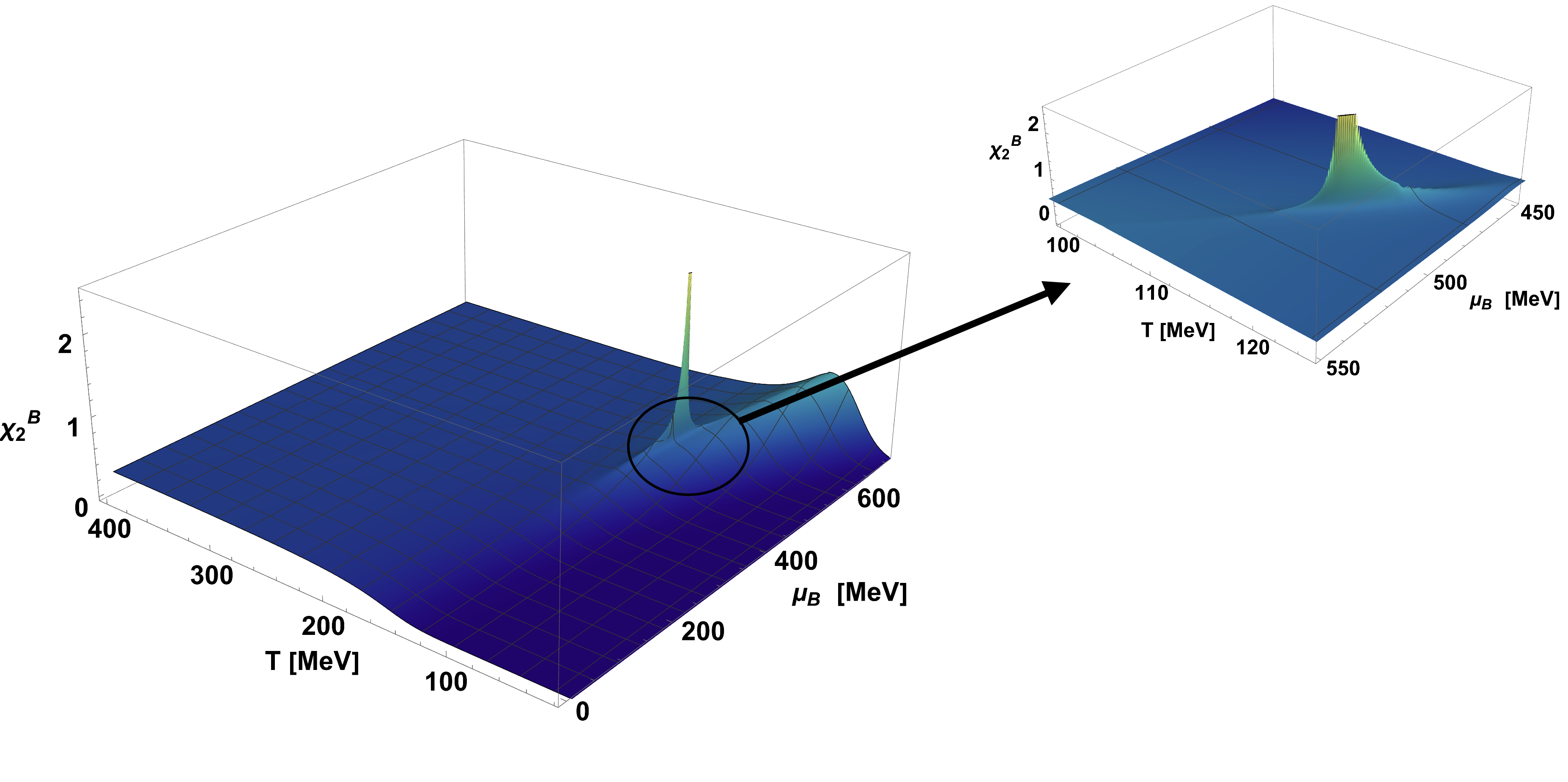}
       \caption{Second baryon number susceptibility as a function of temperature and chemical potential for the same parameters as in Fig.~\ref{fig:Baryondensity2D}, with a zoom into the critical region}
       \label{fig:chi2density3D}
\end{figure*}
\begin{figure*}[!htbp] 
    \includegraphics[width=0.45\textwidth]{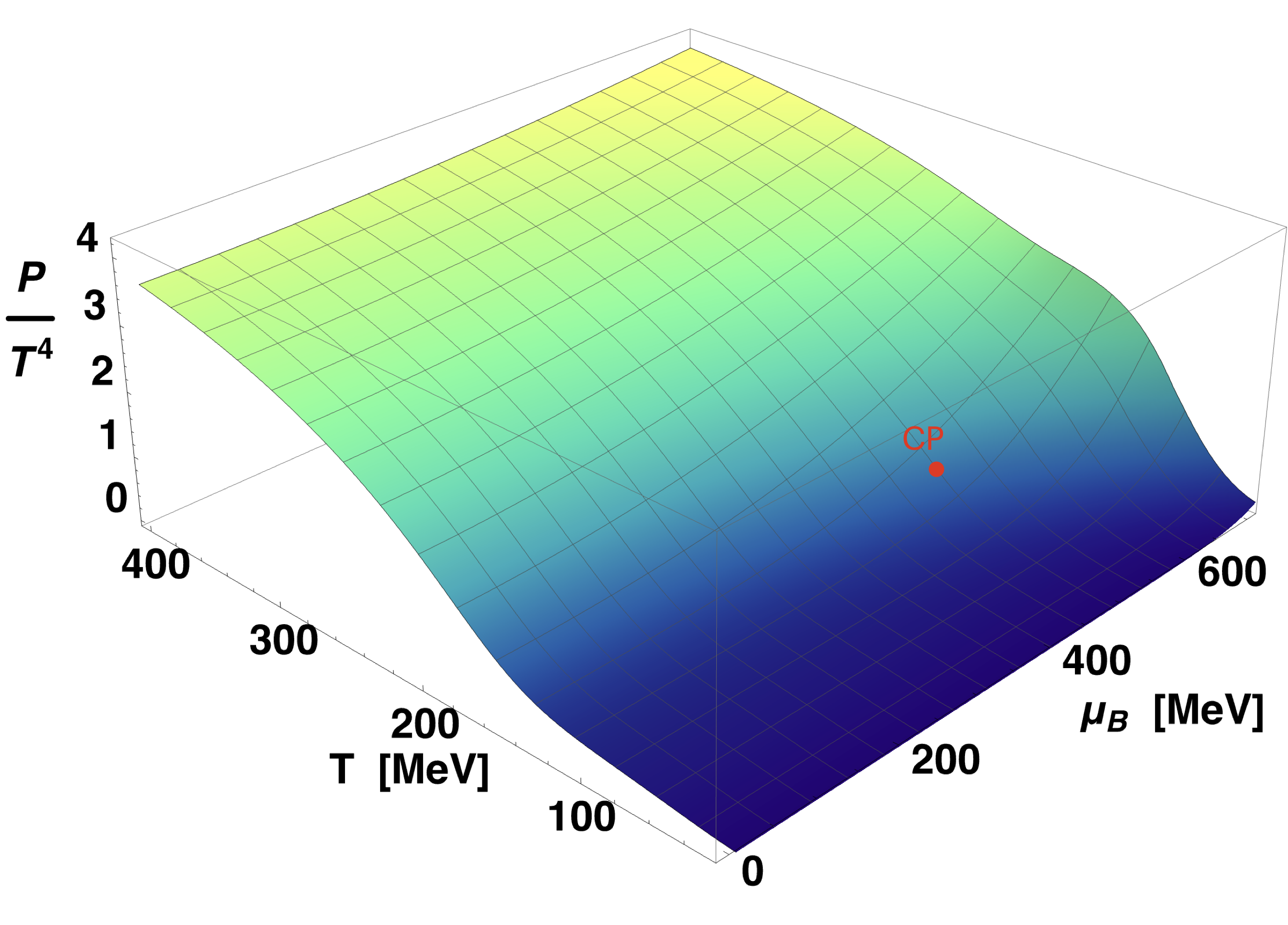}
        \includegraphics[width=0.45\textwidth]{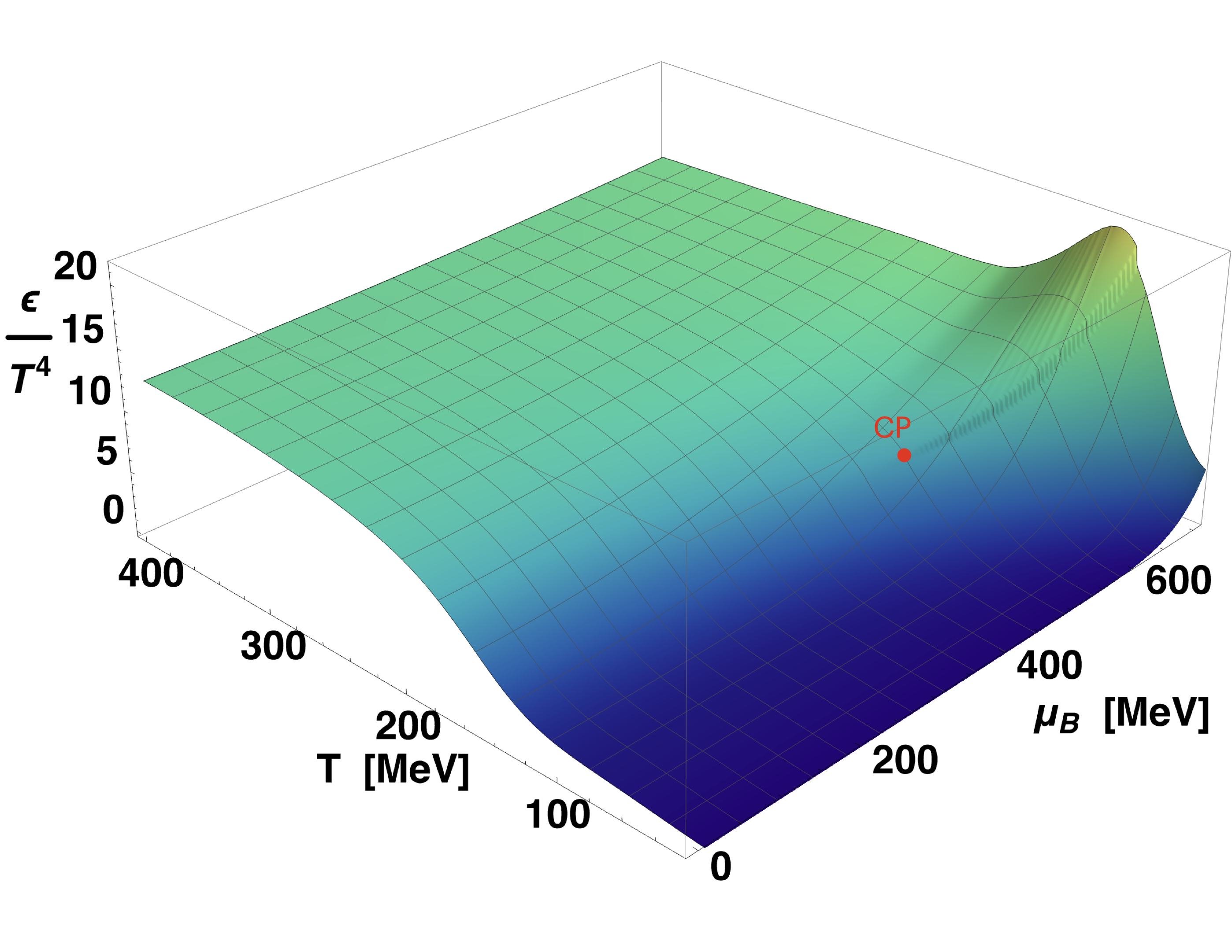}
    \caption{Pressure on the left panel and  Energy density on the right panel with a red point representing a critical point for the same parameters as in Fig.~\ref{fig:Baryondensity2D}}
     \label{fig:Pressuredensity3D}
\end{figure*}

\begin{figure*}[!htbp] 
  \includegraphics[width=0.45\textwidth]{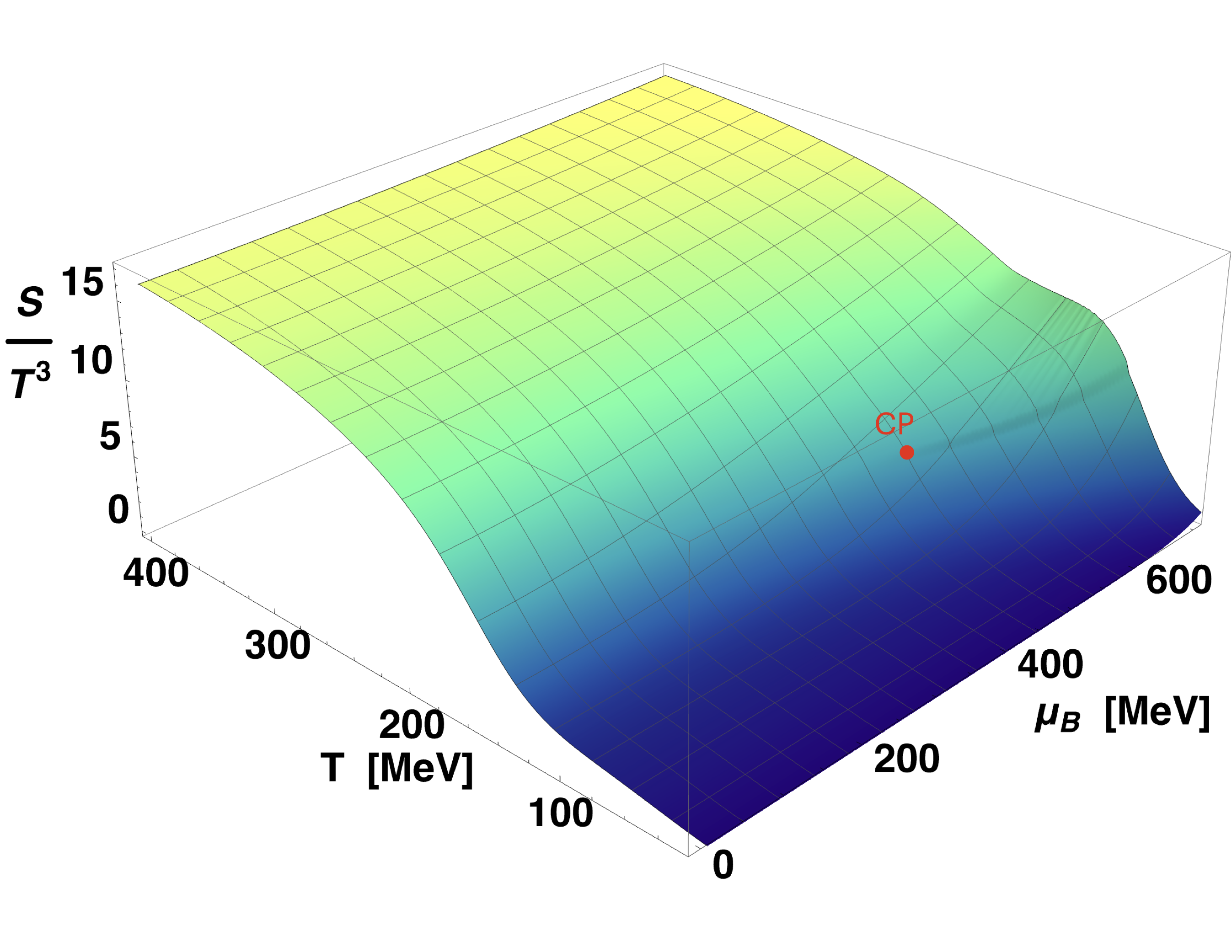}
    \caption{Entropy density as a function of temperature and chemical potential with a red point representing a critical point for the same parameters as in Fig.~\ref{fig:Baryondensity2D}}
    \label{fig:Entropydensity3D}
\end{figure*}

\bibliography{apssamp}

\end{document}